\documentclass[12pt, centerh1]{article}
\usepackage{graphicx}

\usepackage{amsfonts,colonequals,rotating}
\usepackage{natbib, arydshln}
\usepackage{epsfig}
\usepackage{amsfonts} 
\usepackage{amsmath}
\usepackage{amssymb}

\newcommand{\mL}{\mathcal{L}}
\newcommand{\btheta}{{\boldsymbol{\theta}}}
\newcommand{\bvtheta}{{\boldsymbol{\vartheta}}}
\newcommand{\pig}{\pi_g}
\newcommand{\mug}{{\boldsymbol{\mu}}_g}
\newcommand{\nug}{\nu_g}
\newcommand{\Sigg}{{\mathbf{\Sigma}}_g}
\newcommand{\hpig}{\hat{\pi}_g}
\newcommand{\hmug}{\hat{{\boldsymbol{\mu}}}_g}
\newcommand{\hnug}{\hat{\nu}_g}
\newcommand{\hSigg}{\hat{{\mathbf{\Sigma}}}_g}
\newcommand{\hog}{\hat{\omega}_{jg}^{(i)}}
\newcommand{\hzg}{\hat{z}_{jg}^{(i)}}

\newcommand{\bmu}{{\boldsymbol{\mu}}}
\newcommand{\bSig}{{\mathbf{\Sigma}}}
\newcommand{\bLambda}{{\mathbf{\Lambda}}}
\newcommand{\bLambdag}{{\mathbf{\Lambda}_g}}
\newcommand{\bfx}{{\bf x}}

\newcommand{\bfz}{{\bf z}}
\newcommand{\ft}{f_t}
\newcommand{\tsub}{\textsubscript}
\newcommand{\alphset}{{\boldsymbol \alpha}_{\mbox{\tiny ARI}}}
\newcommand\numberthis{\addtocounter{equation}{1}\tag{\theequation}}

\newcommand{\vecmu}{\mbox{\boldmath$\mu$}}

\title{On Fractionally-Supervised Classification: Weight Selection and Extension to the Multivariate $t$-Distribution}
\author{Michael P.B.\ Gallaugher and Paul D.\ McNicholas}
\date{\small Dept.\ of Mathematics \& Statistics, McMaster University, Hamilton, Ontario, Canada.}

\begin{document}

\maketitle{}
\begin{abstract}
Recent work on fractionally-supervised classification (FSC), an approach that allows classification to be carried out with a fractional amount of weight given to the unlabelled points, is further developed in two respects. The primary development addresses a question of fundamental importance over how to choose the amount of weight given to the unlabelled points. The resolution of this matter is essential because it makes FSC more readily applicable to real problems. Interestingly, the resolution of the weight selection problem opens up the possibility of a different approach to model selection in model-based clustering and classification. A secondary development demonstrates that the FSC approach can be effective beyond Gaussian mixture models. To this end, an FSC approach is illustrated using mixtures of multivariate $t$-distributions.

\end{abstract}

\section{Introduction}
In a typical classification application, some of the observations are unlabelled and the objective is to predict the labels of the unlabelled points, for details see \cite{mcnicholas16a}. In such situations, classification is generally semi-supervised or supervised (also called discriminant analysis). These two species of classification differ in whether any weight is given to the unlabelled points in the prediction of their labels. In semi-supervised classification, the labelled and unlabelled points are given equal weight; however, in supervised classification, the unlabelled points are given zero weight. Furthermore, it is possible to either give all the weight to the unlabelled points or treat all the points as unlabelled. This third, and well known, species of classification is called unsupervised classification or cluster analysis. These three species of classification are well established; yet, in any given scenario, it might be the case that labelled or unlabelled observations are more important when building a classifier.

\cite{vrbik15} introduce a general approach, called fractionally-supervised classification (FSC), where classification can be carried out with a fractional amount of weight --- anything between none and all --- being given to the unlabelled points. This approach allows for an intermediate solution between the three different species of classification. Moreover, although it was conceived in the model-based paradigm with the use of Gaussian mixture models and weighted likelihood, discussed in detail in Section~\ref{Background}, it is more generally applicable and will be illustrated herein for $t$-mixtures.

\cite{vrbik15} show that FSC oftentimes improves classification performance when compared to the three different species of classification; however, the problem over how to choose the appropriate amount of weight to give the unlabelled points remains unanswered. \cite{vrbik15} discussed a few different options to choose the appropriate weight but all of these procedures were deemed undesirable. \cite{vrbik15} ultimately decided to use the adjusted Rand index \citep[ARI;][]{hubert85} to choose the weight; however, while this approach was sufficient to illustrate that FSC can be very effective, it is not viable in practice because it assumes knowledge of the labels that are treated as unknown in the analysis. The main contribution of the present work is to determine a weight selection criterion that can be used in real problems, where there are genuinely unlabelled points. The secondary contribution of this paper is the demonstration of FSC for non-Gaussian mixture models, in particular the multivariate $t$-distribution.

The remainder of this paper is laid out as follows. In Section~\ref{Background}, a detailed discussion of mixture models, FSC and weighted likelihood, as well as a brief discussion of the multivariate $t$-distribution is presented. Then, FSC with the multivariate $t$-distribution is laid out (Section~\ref{Method1}) and a detailed discussion on weight selection criteria is presented (Section~\ref{Method2}). In Section~\ref{Analyses}, simulations and demonstrations using real data are presented and we conclude with a discussion and suggestions for future work (Section~\ref{sec:disc}). 


\section{Background}\label{Background}
\subsection{Finite Mixture Models and Model-Based Clustering}
\cite{mcnicholas16a} traces the relationship between mixture models and clustering back as far as \cite{tiedeman55}. The first use of finite mixture models for model-based clustering is generally regarded to be by \cite{wolfe65} and, in the intervening years, model-based clustering has become a popular approach for clustering \citep[a recent review is given by][]{mcnicholas16b}.
 A finite mixture model assumes that an observation $\bfx$ comes from a population with $G$ subgroups. The density function of $\bfx$ is given by
\begin{equation}
f(\bfx\mid{\boldsymbol \vartheta})=\sum_{g=1}^{G}\pig f_{g}(\bfx\mid{\boldsymbol \theta}_g),
\end{equation}
where $\pig>0$, with $\sum_{g=1}^{G}\pig=1$, are called the mixing proportions, $f_g(\cdot)$ are the component densities, and ${\boldsymbol \vartheta}=(\pi_1,\pi_2,\ldots,\pi_G,{\boldsymbol \theta}_1,{\boldsymbol \theta}_2,\ldots,{\boldsymbol \theta}_G)$. 

Because of its mathematical tractability, the Gaussian mixture model has been looked at extensively in the literature. In addition to \cite{wolfe65}, other examples of earlier work in the area of model-based clustering using Gaussian mixtures include \cite{baum70}, \cite{scott71} and \cite{orchard72}. For more details on the history of model based clustering, see \cite{mcnicholas16b}. More recently, there has also been a fair amount of work using non-Gaussian mixtures such as the $t$-distribution \citep[e.g.,][]{peel00, andrews11a, andrews11c, andrews12, steane12, lin14} and skewed distributions \citep{lin10,vrbik12,vrbik14, lee13,lee14, franczak14, franczak15, dang15, lin13, murray14b, murray14a, murray17, murray17b}. Related to this work, an interesting vein of work has been carried out on cluster-weighted models \citep[CWMs; e.g.,][]{ingrassia12,subedi13,ingrassia15,subedi15,punzo17}.

\subsection{Three Species of Classification}
Let the $N\times D$ matrix ${\mathbb X}=(\bfx_1',\bfx_2',\ldots,\bfx_N')'$ be a data matrix, where the $\bfx_i$ are $D$-dimensional vectors and $N$ is the number of data points. We can then split ${\mathbb X}$ into two sub-matrices ${\mathbb X}_1$ and ${\mathbb X}_2$, where ${\mathbb X}_1=(\bfx_{11}',\bfx_{12}',\ldots,\bfx_{1n_1}')'$ are data points with known labels, and ${\mathbb X}_2=(\bfx_{21}',\bfx_{22}',\ldots,\bfx_{2n_2}')'$ are observations with unknown labels. Then write ${\mathbb X}=({\mathbb X}_1,{\mathbb X}_2)'$.
Also, define ${\mathbb Z}=({\mathbb Z}_1,{\mathbb Z}_2)'$, to be a matrix of indicator vectors. Specifically, we define ${\mathbb Z}_1=(\bfz^{(1)'}_1,\bfz^{(1)'}_2,\ldots,\bfz^{(1)'}_{n_1})'$, were $\bfz_{i}^{(1)}$ are $G$-dimensional vectors with elements 0 or 1. For convenience, we will denote element $g$ of $\bfz^{(1)}_j$ by $z_{jg}^{(1)}$ where
$$
z_{jg}^{(1)}=
\left\{
\begin{array}{ll}
1 &\mbox{if } \bfx_{1j} \mbox{ is in group } g,\\
0 & \mbox{otherwise.} 
\end{array}
\right.
$$ 
We can likewise define ${\mathbb Z}_2$ in the same manner. Furthermore, $z_{jg}^{(2)}$ for $j=1 ,2 ,\ldots ,n_2$ are analogous to $z_{jg}^{(1)}$ for the unlabelled observations.
Define $D\tsub{\tiny o}=\{{\mathbb X},{\mathbb Z}_1\}$ to be our set of observed data, and $D\tsub{\tiny C}=\{{\mathbb X},{\mathbb Z}\}$ to be our complete-data. We can furthermore denote the observed data corresponding to labelled observations by $D\tsub{\tiny L}=\{{\mathbb X}_1,{\mathbb Z}_1\}$, and the data corresponding to unlabelled observations by $D\tsub{\tiny U}=\{{\mathbb X}_2\}$. 

Using the above notation, we can now describe the three species of classification. The first species is discriminant analysis, which makes use of only labelled data to build a classifier. The likelihood function in the case of a discriminant analysis can be written as
\begin{equation}
\mathcal{L}\tsub{\tiny DA}(\bvtheta\mid D\tsub{\tiny L})=\prod_{j=1}^{n_1}\prod_{g=1}^{G}\left[\pig f_g(\bfx_{1j}\mid \btheta_g)\right]^{z_{jg}^{(1)}}.
\end{equation}

The second species is cluster analysis, and can take on one of two forms. The first form is the one that we will primarily consider, and makes use of only unlabelled data points and ignores the labelled points. In this case, the likelihood function is given by
\begin{equation}
\mathcal{L}\tsub{\tiny clust}(\bvtheta\mid D\tsub{\tiny U})=\prod_{j=1}^{n_2}\sum_{g=1}^{G}\pig f_g(\bfx_{2j}\mid \btheta_g).
\end{equation}
The second form of the cluster analysis utilizes both labelled and unlabelled points, but treats the labelled points as unlabelled. 

The third species is semi-supervised classification. This makes use of all of the observed data $D\tsub{\tiny o}$ and treats labelled and unlabelled points equally when building a classifier. The likelihood function for semi-supervised classification is given by the product of $\mathcal{L}\tsub{\tiny DA}(\bvtheta\mid D\tsub{\tiny L})$ and $\mathcal{L}\tsub{\tiny clust}(\bvtheta\mid D\tsub{\tiny U})$ to give
\begin{equation}
\mathcal{L}\tsub{\tiny semi}(\bvtheta\mid D\tsub{\tiny o})=\prod_{j=1}^{n_1}\prod_{g=1}^{G}\left[\pig f_g(\bfx_{1j}\mid \btheta_g)\right]^{z_{jg}^{(1)}}\prod_{j=1}^{n_2}\sum_{g=1}^{G}\pig f_g(\bfx_{2j}\mid \btheta_g).
\end{equation}

\subsection{Fractionally-Supervised Classification}\label{FSCG}
Introduced by \cite{vrbik15}, FSC allows for a solution intermediate to the three species of classification. This is achieved by introducing the weight $\alpha_1=\alpha$ to labelled observations, and $\alpha_2=1-\alpha$ to unlabelled observations, where $0\le\alpha\le 1$.  
Using these weights, the most natural form of the weighted observed likelihood can be written as
\begin{equation}\begin{split}\label{eq:orig_likelihood}
\mL_\tsub{\tiny FSC}({\bvtheta}&\mid D\tsub{\tiny o},\alpha)=[\mL\tsub{\tiny DA}({\bvtheta}\mid D\tsub{\tiny L})]^{\alpha}[\mL\tsub{\tiny clust}({\bvtheta}\mid D\tsub{\tiny U})]^{1-\alpha}\\
&=\left[\prod_{j=1}^{n_1}\prod_{g=1}^{G}[\pig f_g(\bfx_{1j}\mid \btheta_g)]^{z_{jg}^{(1)}}\right]^{\alpha}\left[\prod_{j'=1}^{n_2}\sum_{h=1}^{H}\pi_hf_g(\bfx_{2{j'}}\mid \btheta_h)\right]^{1-\alpha},
\end{split}\end{equation}
where $z_{jg}^{(1)}$ is the $g${th} element of $\bfz_{j}^{(1)}$. Although $H$ does not necessarily have to equal $G$, we will make the assumption that $H=G$.
We can then write the complete-data log-likelihood function as
\begin{equation}
\ell(\bvtheta\mid D\tsub{\tiny c}) =\sum_{i=1}^{2}\sum_{j=1}^{n_i}\sum_{g=1}^{G}\alpha_i z_{jg}^{(i)}\left[\log(\pig)+\log(f_g(\bfx_{ij}\mid \btheta_g))\right].
\label{eq:cwll}
\end{equation}

The expectation-maximization (EM) algorithm \citep{dempster77} can then be used to maximize \eqref{eq:cwll}. The EM algorithm is an iterative algorithm that, on each iteration, consists of a conditional expectation (E-) step and the subsequent maximization of the expectation (M-step). We first initialize the parameters, and we denote this by $\bvtheta^{(0)}$. Iteration $t+1$ of the EM algorithm proceeds as follows.
\begin{subequations}
\begin{align}
&\mbox{{\bf E-Step:} Calculate }  Q\big(\bvtheta\mid \bvtheta^{(t)}\big)={\mathbb E}_{{\mathbb Z}_2\mid {\mathbb X}}\big[\ell\left(\bvtheta\mid D\tsub{\tiny c}\right)\mid D\tsub{\tiny o},\bvtheta^{(t)}\big] \label{EM1}\\
&\mbox{{\bf M-Step:} Find } \underset{\bvtheta}\arg\max \hspace{0.1cm} Q\big(\bvtheta\mid \bvtheta^{(t)}\big) \label{EM2}\\
&\mbox{Check for convergence. If the convergence criterion was not met, set } t=t+1\\\nonumber 
& \mbox{and return to \eqref{EM1}.} \label{EM3}
\end{align}
\end{subequations}
As was shown in \cite{vrbik15}, in the case of a Gaussian mixture model, steps \eqref{EM1} and \eqref{EM2} simplify to the following.
\begin{subequations}
\begin{align*}
&\mbox{{\bf E-Step:} Update }\\ 
&\qquad\qquad\qquad\qquad\hat{z}_{jg}^{(2)}=\frac{\pig^{(t)}\phi(x_{2j}\mid \mug^{(t)},\Sigg^{(t)})}{\sum_{g=1}^{G}\pig^{(t)}\phi(x_{2j}\mid \mug^{(t)},\Sigg^{(t)})}.\\
&\mbox{Because the } z_{jg}^{(1)} \mbox{ are known, we set } \hat{z}_{jg}^{(1)}=z_{jg}^{(1)}.\\
&\mbox{{\bf M-Step:} Update the estimates of $\pig$, $\mug$ and $\Sigg$ by calculating}\\ 
&\qquad\qquad\qquad\qquad\pig^{(t+1)}=\frac{S_g}{\sum_{g=1}^G S_g},\quad
\mug^{(t+1)}=\frac{\sum_{i=1}^2\sum_{j=1}^{n_i}\alpha_i\hat{z}_{jg}^{(i)}\bfx_{ij}}{S_g},\\
&\qquad\qquad\qquad\qquad\Sigg^{(t+1)}=\frac{\sum_{i=1}^{2}\sum_{j=1}^{n_i}\alpha_i\hat{z}_{jg}^{(i)}(\bfx_{jg}-\mug^{(t+1)})(\bfx_{jg}-\mug^{(t+1)})'}{S_g},\\
&\mbox{where } S_g=\sum_{i=1}^2\sum_{j=1}^{n_i}\alpha_i\hat{z}_{jg}^{(i)}.
\end{align*}
\end{subequations}
This simplified form of the EM algorithm will prove useful when we discuss the EM algorithm in the case of a FSC with mixture of multivariate $t$-distributions (Section~\ref{Method1}).

We note that the three different species of classification fall out naturally as special cases of FSC. If $\alpha=1$, then all of the weight is given to the labelled observations, and the unlabelled observations are ignored. In this case, we are performing discriminant analysis. If $\alpha=0.5$, then the labelled and unlabelled observations are given equal weight, and we are then performing semi-supervised classification. Finally, if $\alpha=0$, then no weight is given to the labelled observations, and thus we are performing a cluster analysis (on the unlabelled observations). 
As mentioned in Section~1, the main unresolved issue with FSC is the selection of the weight $\alpha$. 

\subsection{The Multivariate $t$-Distribution}
The $p$-dimensional $t$-distribution with $\nu$ degrees of freedom, location parameter $\bmu$ and scale matrix $\bSig$, arises as a special case of a normal scale mixture \citep{peel00}. Specifically, we can write the normal scale mixture as
\begin{equation}
\epsilon\phi(\bfx\mid \bmu, \bSig)+(1-\epsilon)\phi(\bfx\mid \bmu, \nu\bSig),
\label{eq:NSM}
\end{equation}
where $\phi(\cdot)$ denotes the multivariate Gaussian density with mean $\bmu$ and covariance matrix $\bSig$, and $\epsilon$ is small. We can then rewrite \eqref{eq:NSM} as
$$
\int\phi(\bfx\mid \bmu,\nu\bSig)dH(w),
$$ 
where
\begin{equation}
H(w)=\frac{1}{\Gamma\left(\frac{\nu}{2}\right)\left(\frac{\nu}{2}\right)^{\frac{\nu}{2}}}w^{\frac{\nu}{2}-1}\exp\left\{-\frac{2w}{\nu}\right\},
\label{gammapdf}
\end{equation}
$w>0$ and $\Gamma(\cdot)$ is the gamma function. Note that \eqref{gammapdf} is the probability density function of a gamma$({\nu}/{2},{\nu}/{2})$ random variable. 
The resulting density for the multivariate $t$-distribution is
\begin{equation}
f_t(\bfx\mid \bmu,\bSig,\nu)=\frac{\Gamma\left(\frac{\nu+p}{2}\right)|\bSig|^{-\frac{1}{2}}}{(\pi\nu)^{\frac{1}{2}p}\Gamma\left(\frac{\nu}{2}\right)\left[1+\frac{\delta(\bfx,\bmu,\bSig)}{\nu}\right]^{\frac{1}{2}(\nu+p)}},
\label{eq:tdist}
\end{equation}
where $\delta(\bfx,\bmu,\bSig)=(\bfx-\bmu)'\bSig^{-1}(\bfx-\bmu)$ is the squared Mahalanobis distance.
Maximum likelihood estimation for $t$-mixtures, in the context of model-based clustering, utilizes the introduction of latent variables $W_{ig}$ such that
$
W_{ig}\mid z_{ig}=1\sim\mbox{gamma}\left({\nu_g}/{2},{\nu_g}/{2}\right).
$ 
\subsection{Parsimonious Models}
The eigen-decomposition of a matrix is widely used in both mathematics and multivariate statistics. In the context of mixture models, we can write a covariance, or scale, matrix in the form
$
\Sigg=\lambda_g\bLambdag{\bf D}_g\bLambdag',
$
where $\lambda_g$ is a constant, ${\bf D}_g$ is a diagonal matrix with entries that are proportional to the eigenvalues, and $\bLambdag$ is a matrix of eigenvectors. We can then impose the following constraints: 
$\lambda_g=\lambda, \ \bLambdag=\bLambda, \ \bLambdag={\bf I}, \ {\bf D}_g={\bf D}, \ {\bf D}_g={\bf I},$
where ${\bf I}$ is the identity matrix \citep{banfield93,celeux95}. \cite{celeux95} employ combinations of the above constraints to the covariance matrices in a Gaussian mixture model to form a family of 14 Gaussian parsimonious clustering models (GPCMs). Of these 14 models, 12.   
are extended to the $t$-distribution by \cite{andrews12}, with the result known as the tEIGEN family. 
These 12 models, together with the option to constrain $\nu_g=\nu$, leads to 24 different models in the tEIGEN family. The current form of the tEIGEN package \citep{andrews15} in {\sf R} \citep{R15} supports all 14 GPCM scale structures and hence a family of 28 tEIGEN models, which are summarized in Table \ref{tab:teigen} (Appendix~\ref{app:tEIGEN}).

\subsection{Model Selection Criteria}
We now discuss a couple of criteria that are commonly used to select an appropriate parsimonious model. The most common approach is the Bayesian information criterion \citep[BIC;][]{schwarz78}, 
which is given by
$$
\text{BIC}=2\ell\tsub{\tiny obs}(\bvtheta\mid D\tsub{\tiny o})-p\log N,
$$
where $\ell\tsub{obs}$ is the maximized observed likelihood, $p$ is the number of free parameters, and $N$ is the total number of data points. The BIC has been frequently used for parsimonious model selection, e.g., \cite{fraley98} and \cite{mcnicholas08}.
Another criterion that is widely used is the integrated completed likelihood \citep[ICL;][]{biernacki00}, which penalizes the BIC for classification uncertainty. The ICL can be approximated using the BIC:
$$
\text{ICL}\approx\text{BIC}-2\sum_{i=1}^{n_g}\sum_{g=1}^G\text{MAP}(\hat{z}_{ig})\log\hat{z}_{ig},
$$
where
$$
\text{MAP}(\hat{z}_{ig})=
\left\{
\begin{array}{ll}
1 & \mbox{if } \arg\max_{h=1,\ldots,G}\{\hat{z}_{ih}\}=g,\\
0 & \mbox{otherwise.}
\end{array}
\right.
$$


\section{FSC for $t$-Mixtures}\label{Method1}
Before we discuss FSC for $t$-mixtures, we note that there is an alternative form of the weighted likelihood; for completeness, this is discussed in Appendix~\ref{app:b}. The main complication when using $t$-mixtures, compared to using Gaussian mixtures, is the update for the degrees of freedom. This update, unfortunately, has no closed form and has to be calculated using numerical methods.
The incomplete weighted observed likelihood when using multivariate $t$ component densities is
$$
\mL\tsub{obs}(\bvtheta\mid D\tsub{\tiny o},\alpha)=\left[\prod_{j=1}^{n_1}\prod_{g=1}^{G}[\pig\ft(\bfx_{1j}\mid\mug,\Sigg,\nug)]\right]^{\alpha}\left[\prod_{j'=1}^{n_2}\sum_{g=1}^{G}\pig\ft(\bfx_{2{j'}}\mid\mug,\Sigg)\right]^{1-\alpha},
$$
where $\ft(\cdot)$ is the density for the multivariate $t$-distribution defined in \eqref{eq:tdist}. To find $\arg\max_{\btheta}\mL\tsub{obs}$, we use a multicycle ECM algorithm similar to \cite{andrews12}. After initializing $z_{jg}^{(i)}$ and $w_{jg}^{(i)}$, iteration $t+1$ of the multicycle ECM algorithm would proceed as follows:
\begin{subequations}
\begin{align}
&\mbox{{\bf E-Step}: Update}\nonumber\\ 
&\qquad\qquad\hat{z}_{jg}^{(2)}=\frac{\hpig\ft(\bfx_{2j}\mid\hmug^{(t)},\hSigg^{(t)},\hnug^{(t)})}{\sum \limits_{g=1}^{G}\hpig\ft(\bfx_{2j}\mid\hmug^{(t)},\hSigg^{(t)},\hnug^{(t)})},
\label{eq:tEstepz}\\
&\qquad\qquad\hat{w}_{jg}^{(i)}=\frac{\hnug^{(t)}+p}{\hnug^{(t)}+\delta(\bfx_{ij},\hmug^{(t)},\hSigg^{(t)},\hnug^{(t)})}.
 \label{eq:tEstepw}\\
&\mbox{{\bf First CM-Step}: Update } \hpig, \hmug \mbox{ and } \hnug. \mbox{ The updates for $\hpig$, and $\hmug$ are given in}\nonumber\\
&\mbox{closed form as}\nonumber
\end{align}
\end{subequations}
\begin{subequations}
\begin{align}
&\qquad\qquad\hpig^{(t+1)}=\frac{\sum \limits_{i=1}^{2}\sum \limits_{j=1}^{n_i} \alpha_{i}\hat{z}_{jg}^{(i)}}{\sum \limits_{i=1}^{2}\sum \limits_{j=1}^{n_i} \sum \limits_{g=1}^{G} \alpha_{i}\hat{z}_{jg}^{(i)}} 
\quad\mbox{ and }\quad
\hmug^{(t+1)}=\frac{\sum \limits_{i=1}^{2}\sum \limits_{j=1}^{n_i} \alpha_{i}\hat{z}_{jg}^{(i)}\hat{w}_{jg}^{(i)}\bfx_{ij}}{\sum \limits_{i=1}^{2}\sum \limits_{j=1}^{n_i} \alpha_{i}\hat{z}_{jg}^{(i)}\hat{w}_{jg}^{(i)}}.\nonumber\\
&\mbox{The updates for the degrees of freedom } \nug, \mbox {as mentioned before, do not have a closed}\nonumber\\ &\mbox{form and have to be calculated using numerical methods. In the unconstrained case,}\nonumber\\ &\mbox{one has to solve \eqref{eq:dfup} for } \hnug^{\text{\tiny new}}.\nonumber\\ 
&\qquad\qquad-\varPsi\left(\frac{1}{2}\hnug^{\text{\tiny new}}\right)+\log\left(\frac{1}{2}\hnug^{\text{\tiny new}}\right)-\varPsi\left(\frac{\hnug+p}{2}\right)-\log\left(\frac{\hnug+p}{2}\right)+1\nonumber\\
&\qquad\qquad\qquad\qquad+\frac{1}{m_g}\sum \limits_{i=1}^{2} \sum \limits_{j=1}^{n_i} \alpha_i\hat{z}_{jg}^{(i)}\left(\log\hog-\hog\right)=0
\label{eq:dfup}\\
&\mbox{where}\nonumber\\ 
&\qquad\qquad m_g=\sum \limits_{i=1}^{2}\sum \limits_{j=1}^{n_i}\alpha_i\hat{z}_{jg}^{(i)}\nonumber\\
&\mbox{and } \varPsi(\cdot) \mbox{ is the digamma function. Then, set } \hnug^{(t+1)}=\hnug^{\text{\tiny new}}. \mbox{ Note that we used the}\nonumber\\ & \texttt{uniroot } \mbox{function in {\sf R} to solve \eqref{eq:dfup}.}\nonumber\\
&\mbox{{\bf E-Step:} Update } \hat{z}_{jg}^{(2)} \mbox{ and } \hat{w}_{jg}^{(i)} \mbox{ using \eqref{eq:tEstepz} and \eqref{eq:tEstepw} with current parameter estimates.}\nonumber\\
&\mbox{{\bf Second CM Step}: Update } \Sigg. \text{ In the completely unconstrained case, the update is}\nonumber\\
&\qquad\qquad \hSigg^{(t+1)}=\frac{1}{m_g}\sum \limits_{i=1}^{2}\sum\limits_{j=1}^{n_i}\sum\limits_{g=1}^{G}\alpha_i\hzg\hog(\bfx_{ij}-\hmug^{(t+1)})(\bfx_{ij}-\hmug^{(t+1)})'.\nonumber
\end{align}
\end{subequations}
We take this time to note that, except for the inclusion of the weights, the multicycle ECM algorithm described here is exactly the same as that described in \cite{andrews12}.

We perform k-means clustering \citep{macqueen67} with 50 random starts to initialize the ECM algorithm, and the Aitken acceleration \citep{aitken26} procedure described in \cite{mcnicholas10a} as our convergence criteria.
Because of the updates for the degrees of freedom, fitting FSC with a $t$-mixture becomes more computationally expensive than fitting a Gaussian model. However, because of the heavier tails of the $t$-distribution, the $t$-mixture is more robust to outlying observations.

\section{Weight Selection Criteria}\label{Method2}
The ARI compares two different partitions of a dataset and, in the classification paradigm, a value of 1 corresponds to perfect classification, whereas a value of 0 indicates that the classification solution is as would be expected if the labels were randomly assigned. In Section~1, we point out that \cite{vrbik15} use the ARI as a weight selection criteria for FSC. However, this is only useful when exploring the overall performance of FSC in simulations and datasets where all the labels are known (but some are treated as unknown). In a real classification scenario, not all the labels will be known and hence the ARI could not be used to select the weight $\alpha$. We, therefore, try other criteria for weight selection.

The first criteria we consider are the BIC and ICL. The results are not shown here but suffice it to say that various analyses revealed both of these criteria to be monotone in $\alpha$ and a boundary point was always chosen. Three different classification-based criteria are considered: the entropy, an alternative form of the entropy \citep{celeux96},  and the $U$ criterion \citep{bensmail97}. 

In our case, the entropy $E$ can be written
\begin{equation}\label{eqn:entropy}
E=\sum_{i=1}^{2}\sum_{j=1}^{n_i}\sum_{g=1}^{G}\text{MAP}(\hat{z}_{jg}^{(i)})\log \hat{z}_{jg}^{(i)}= \sum_{j=1}^{n_2}\sum_{g=1}^{G}\text{MAP}(\hat{z}_{jg}^{(2)})\log \hat{z}_{jg}^{(2)},
\end{equation}
where 
$$
\text{MAP}(z_{jg}^{(i)})=
\left\{
\begin{array}{ll}
1 & \mbox{if } \hat{z}_{jg}^{(i)}=\max_{h=1,2,\ldots,G}(\hat{z}_{jh}^{(i)}),\\
0& \mbox{otherwise},
\end{array}
\right.
$$
and taking $0\log 0=0$. The entropy in \eqref{eqn:entropy} is always negative, unless there is no uncertainty in the clustering solution, in which case it is $0$. When using this criterion, we choose the optimal weight to correspond to the maximum value of $E$. 

An alternative form of the entropy is sometimes used that eliminates the MAP. The resulting criterion, in our case, is given by
$$
A=\sum_{j=1}^{n_2}\sum_{g=1}^{G}\hat{z}_{jg}^{(2)}\log \hat{z}_{jg}^{(2)}.
$$ 
Once again, we choose the optimal weight to correspond to the maximum value of $A$. 
The third, and final, classification-based criterion that we consider is the $U$ criterion. In our case, this is given by
\begin{align*}
U&=\sum_{i=1}^{2}\sum_{j=1}^{n_i}\min_{g=1,2,\ldots,G}(1-\hat{z}_{jg}^{(i)})
=\sum_{j=1}^{n_2}\min_{g=1,2,\ldots,G}(1-\hat{z}_{jg}^{(2)}).
\end{align*}
We observe that $U$ is always positive and, if there is no uncertainty in the classification solution, then $U=0$. Again, we choose the optimal weight to correspond to the maximum value of $U$. 

In addition to these three classification-based criteria, we consider two non-parametric criteria. Before the BIC became popular, the sum of squares matrix was used as a basis for criteria to choose the number of groups in a model \cite[see][Sec.~3.3, for discussion]{gordon81}. Assuming that our data matrix ${\bf X}$ has been partitioned into $G$ groups, we can define the total sum of squares matrix to be
$$
{\bf S}=\sum_{i=1}^{n_g}\sum_{g=1}^{G}(\bfx_{ig}-\bar{\bfx}_g)(\bfx_{ig}-\bar{\bfx}_g)'.
$$
Using a decomposition of ${\bf S}$ we can write
$$
{\bf S}={\bf W}+{\bf B},
$$
where ${\bf W}$ is the within cluster sum of square matrix defined as
$$
{\bf W}=\sum_{g=1}^{G}\sum_{i=1}^{n_g}(\bfx_{ig}-\bar{\bfx}_{g})(\bfx_{ig}-\bar{\bfx}_{g})',
$$
where $\bar{\bfx}_g$ is the sample mean of group $g$, and {\bf B} is the between cluster sum of squares matrix defined as
$$
{\bf B}=\sum_{g=1}^{G}(\bar{\bfx}_g-\bar{\bfx})(\bar{\bfx}_g-\bar{\bfx})',
$$
where $\bar{\bfx}$ is the grand mean.
Although the principle of using the sum of squares matrix was considered all the way back in the 1960s \citep[e.g.,][]{edwards65,friedman67}, it is still visible within the modern literature \cite[e.g.,][]{andrews14}. Herein (Section~\ref{sec:weightex}), two different criteria that use the within cluster sum of squares matrix ${\bf W}$ are tried. The first criterion is based on minimizing the trace of $\mathbf{W}$, i.e., $\text{tr}(\mathbf{W})$, and the second criterion is based on minimizing the determinant of $\mathbf{W}$, i.e., $\text{det}(\mathbf{W})$. 

\section{Analyses}\label{Analyses}

\subsection{Specifying the Number of Groups}
For the purposes of our simulations and data analyses, we assume that the number of groups is equal to the number of components or classes present in the labelled points. However, this could be potentially problematic. For one, there could be a group present in the population that is not represented in the labelled data --- this may be more likely if only a small proportion of the data points are labelled. Although perhaps less likely, it is also possible for the true number of groups to be less than that indicated by the labels. The former problem can be handled by fitting FSC with a different number of groups $H\ge G$ in the cluster analysis component of the likelihood, and then using a criterion such as the BIC or ICL to choose the number of groups. The latter case, however, would need to be treated more carefully; likely in conjunction with a subject matter (data) expert.

\subsection{Simulations}\label{sec:sims}
\begin{figure}[!ht]
\centering
\includegraphics[width=0.6\textwidth]{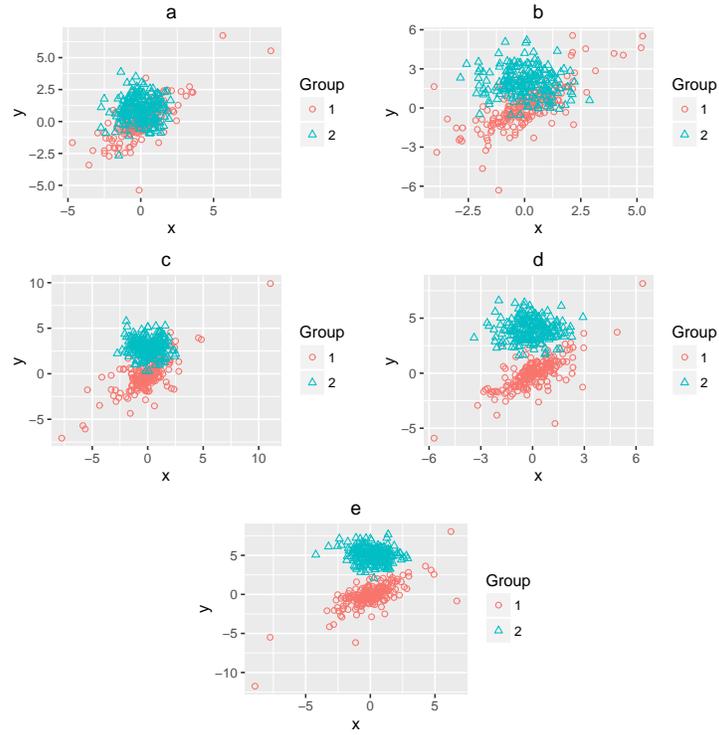}
\caption{Typical datasets for (a) $\Delta=1$,(b) $\Delta=2$,(c) $\Delta=3$,(d) $\Delta=4$,(e) $\Delta=5$,.}
\label{fig:Data}
\end{figure}

Simulations are performed, similar to those in \cite{vrbik15}, to demonstrate FSC with the multivariate $t$-distribution. In all, 100 datasets are simulated, each with 200 data points and two groups. The first group follows a $t_2({\bf 0},{\mathbf\Sigma}_1,\nu_1)$ distribution, where $\nu_1=3$, and 
$$
{\mathbf \Sigma}_1=
\left[
\begin{array}{cc}
1&0.7\\
0.7&1\\
\end{array}
\right].
$$
The second group is taken from a $t_2({\mathbf \Delta},{\mathbf \Sigma}_2,\nu_2)$ distribution, where ${\mathbf \Delta}=[0,\Delta]'$, $\nu_2=70$, and 
$$
{\mathbf \Sigma}_2=
\left[
\begin{array}{cc}
1&0\\
0&1\\
\end{array}
\right].
$$
In this case, one group has a multivariate $t$-distribution, while the other group is approximately Gaussian, i.e., $\nu_2$ is quite large. This time, we take $\Delta\in\{1,2,3,4,5\}$ and the same percentages of labelled data $p$ as previously. In \figurename~\ref{fig:Data}, we show example datasets for each $\Delta$. 

To choose the weights for FSC, we consider 11 candidate values of $\alpha$; specifically, $\alpha\in{\boldsymbol \alpha}_{\tiny\mbox{ARI}}$, where ${\boldsymbol \alpha}_{\tiny\mbox{ARI}}=\{0,0.1,0.2,\ldots,1\}$. Then, the ARI is calculated for each of these weights for the 100 datasets and the average ARI is computed for each weight. The weight with the highest average ARI is chosen. The resulting FSC solution for each weight $\alpha$ is denoted by FSC\tsub{$\alpha$}. Furthermore, for the FSC solution with the chosen weight resulting from the highest average ARI, the notation FSC\tsub{\tiny ARI} is used. Finally, in the special cases corresponding to the three species of classification $\alpha=0,0.5,1$, the FSC solution is denoted by FSC\tsub{\tiny clust}, FSC\tsub{\tiny class} and FSC\tsub{\tiny DA}, respectively. 

\begin{figure}[!htb]
\centering
\includegraphics[width=0.62\textwidth]{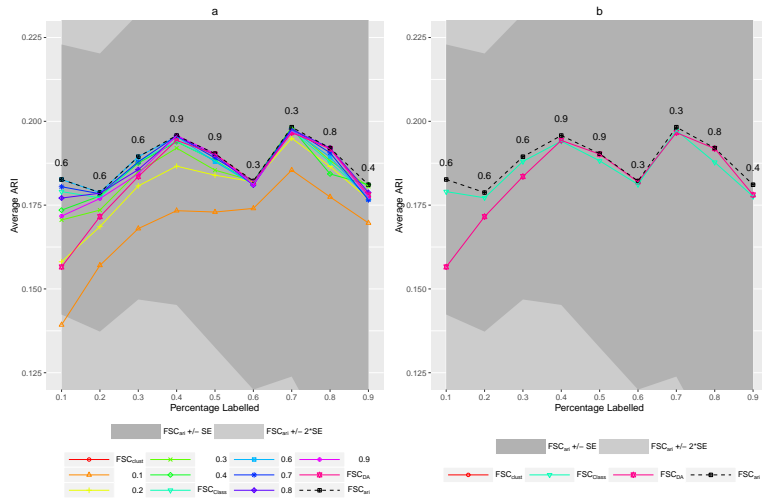}
\caption[$t$-Distribution Results for $\Delta=1$]{For $\Delta=1$: a) FSC\tsub{$\alpha$} and FSC\tsub{\tiny ARI} for $\alpha\in\alphset$, b) FSC\tsub{\tiny clust},FSC\tsub{\tiny class}, FSC\tsub{\tiny DA} and FSC\tsub{\tiny ARI}.}
\label{fig:FSCTd1}
\end{figure}

In \figurename~\ref{fig:FSCTd1}, we give line plots for the case when $\Delta=1$, where the average ARI is plotted against the percentage of labelled data $p$ for each candidate weight. Further, a black dotted line is used to show the result for FSC\tsub{\tiny ARI} with the corresponding chosen weight shown above each point. The left plot shows the results when using all the weights and the right plot singles out the three different species of classification and FSC\tsub{\tiny ARI}. The standard errors are calculated by taking the ARI for all 100 datasets of the chosen weight of FSC\tsub{\tiny ARI} and calculating one (darker grey) and two (lighter grey) standard deviations from the mean ARI. 

\begin{figure}[!htb]
\centering
\includegraphics[width=0.62\textwidth]{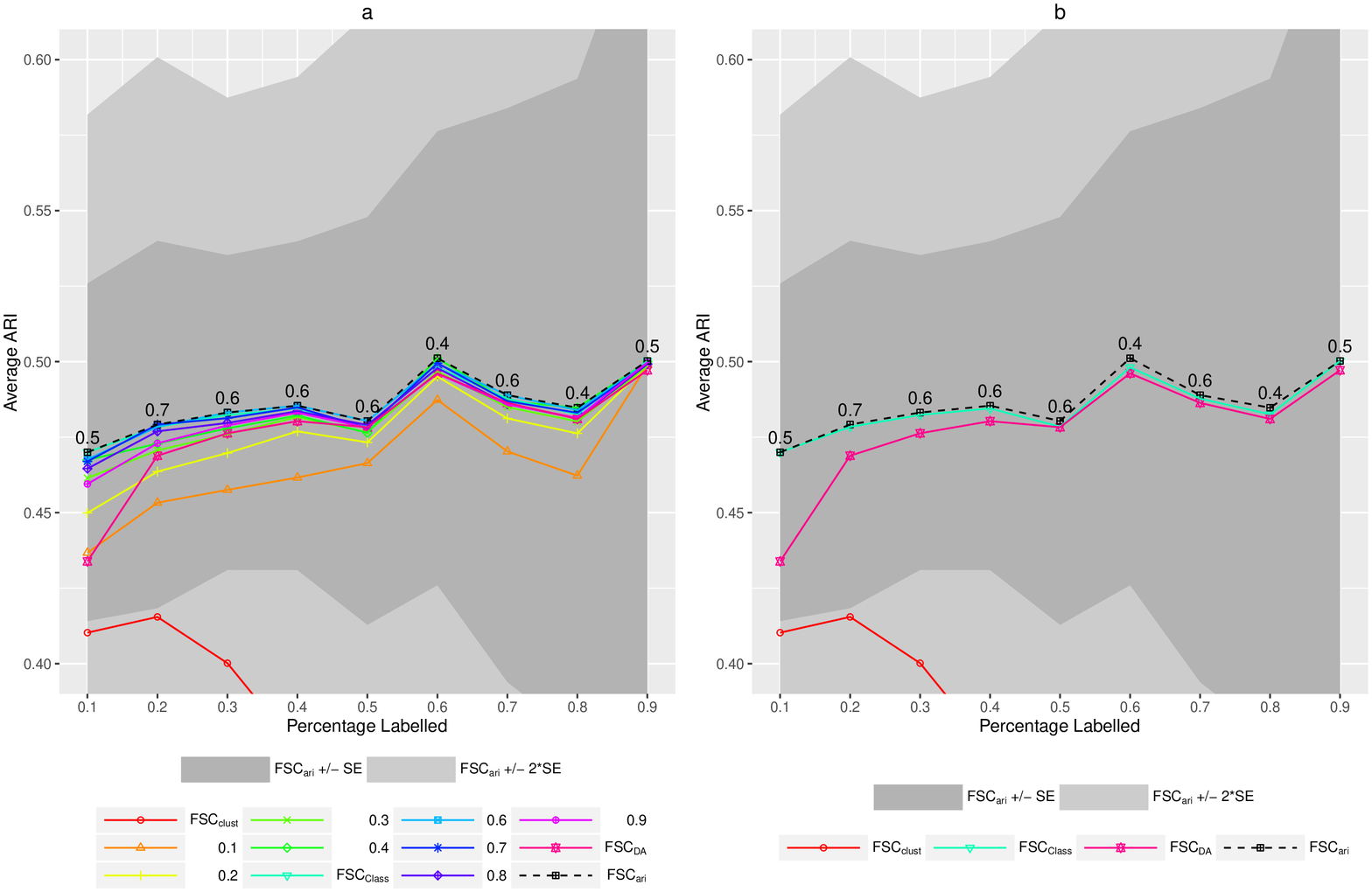}
\caption[$t$-Distribution Results for $\Delta=2$]{For $\Delta=2$: a) FSC\tsub{$\alpha$} and FSC\tsub{\tiny ARI} for $\alpha\in\alphset$, b) FSC\tsub{\tiny clust},FSC\tsub{\tiny class}, FSC\tsub{\tiny DA} and FSC\tsub{\tiny ARI}.}
\label{fig:FSCTd2}
\end{figure}

\begin{figure}[!thb]
\centering
\includegraphics[width=0.62\textwidth]{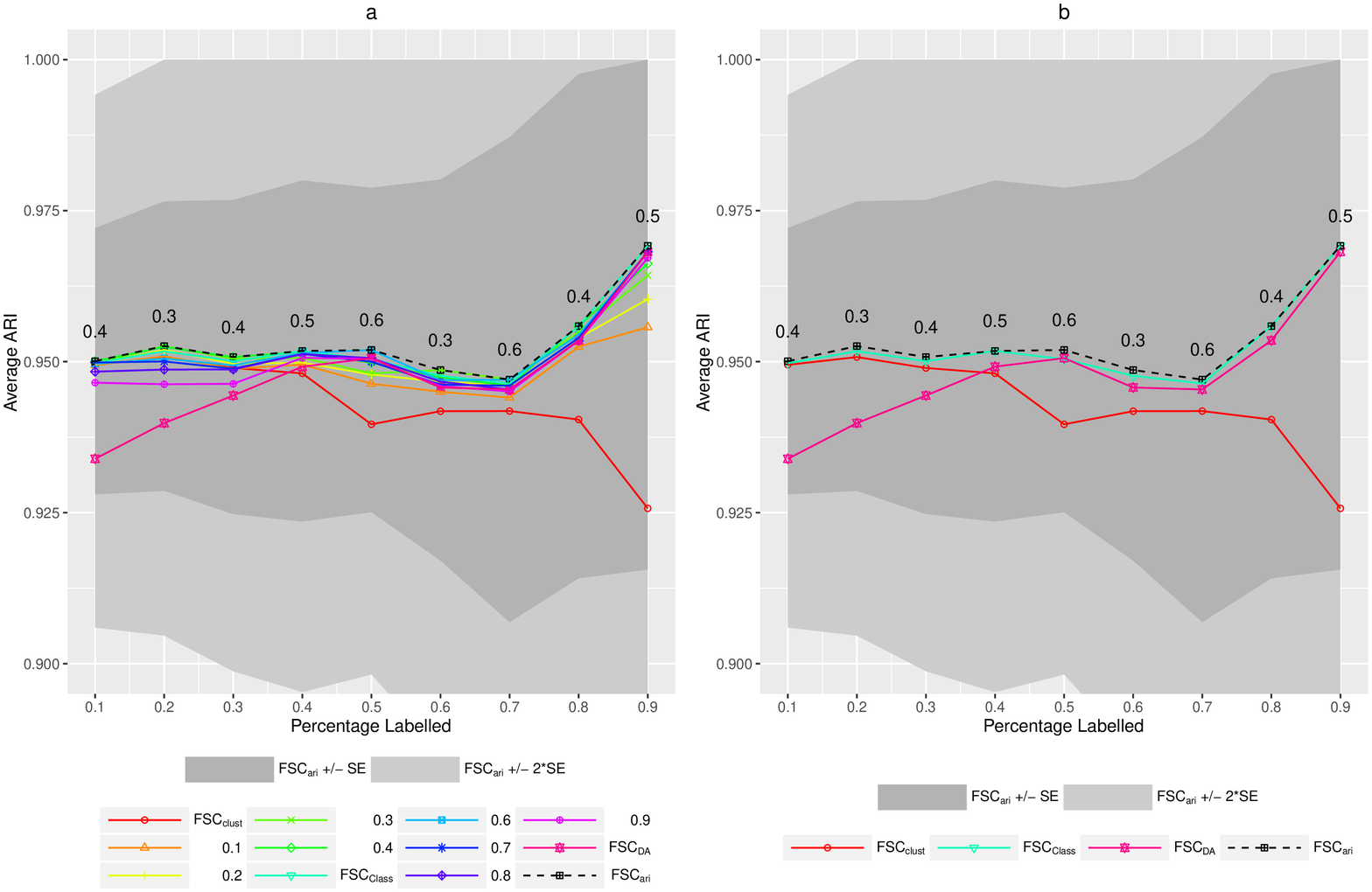}
\caption[$t$-Distribution Results for $\Delta=5$]{For $\Delta=5$: a) FSC\tsub{$\alpha$} and FSC\tsub{\tiny ARI} for $\alpha\in\alphset$, b) FSC\tsub{\tiny clust},FSC\tsub{\tiny class}, FSC\tsub{\tiny DA} and FSC\tsub{\tiny ARI}.}
\label{fig:FSCTd5}
\end{figure}

For $\Delta=1$, we notice that the line for FSC\tsub{\tiny clust} does not appear because the average ARI for each percentage of labelled data is quite small in comparison to the other weights (see \figurename~\ref{fig:FSCTd1}). Furthermore, for all other values of $\Delta$, FSC\tsub{\tiny clust} has the worst performance at higher percentages of labelled data, which is somewhat expected. We also see that all of the chosen weights correspond to a non-species solution. Furthermore, it is interesting to point out that, for lower percentages of labelled data, more weight is given to the labelled points and, at higher percentages, with the exception of $80$\%, less weight is given to the labelled observations. Similar results are given in \figurename~\ref{fig:FSCTd2}--\ref{fig:FSCTd5}, where similar plots are shown for the other values of $\Delta$. 
For the remaining values of $\Delta$, of the 36 different cases, the chosen weight corresponds to a species of classification only nine times. Of these nine occurrences, eight of them correspond to semi-supervised classification, one corresponds to a discriminant analysis, and none of them correspond to a cluster analysis. 
\begin{figure}[!htb]
\centering
\includegraphics[width=0.62\textwidth]{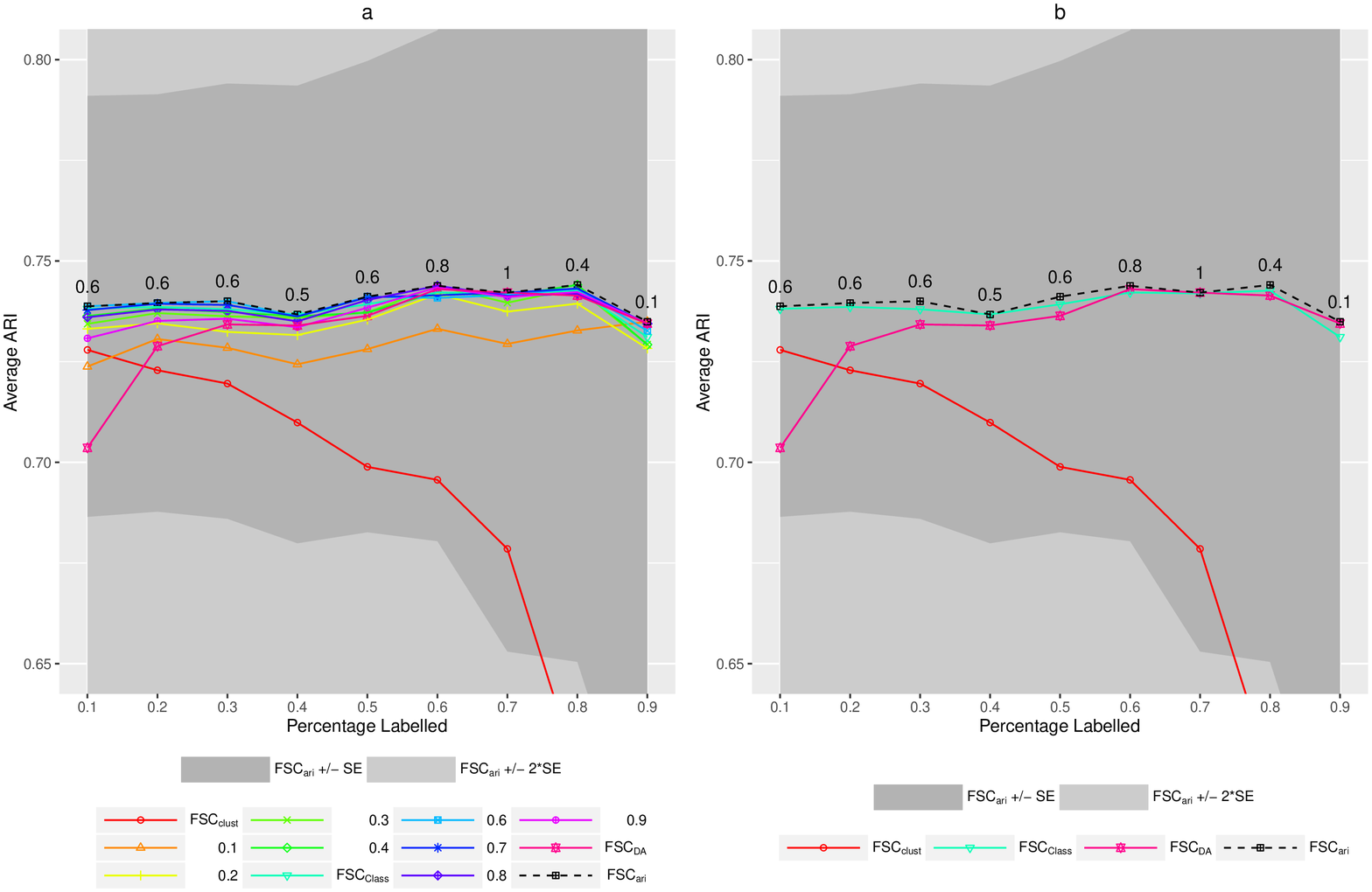}
\caption[$t$-Distribution Results for $\Delta=3$]{For $\Delta=3$: a) FSC\tsub{$\alpha$} and FSC\tsub{\tiny ARI} for $\alpha\in\alphset$, b) FSC\tsub{\tiny clust},FSC\tsub{\tiny class}, FSC\tsub{\tiny DA} and FSC\tsub{\tiny ARI}.}
\end{figure}
\begin{figure}[!htb]
\centering
\includegraphics[width=0.62\textwidth]{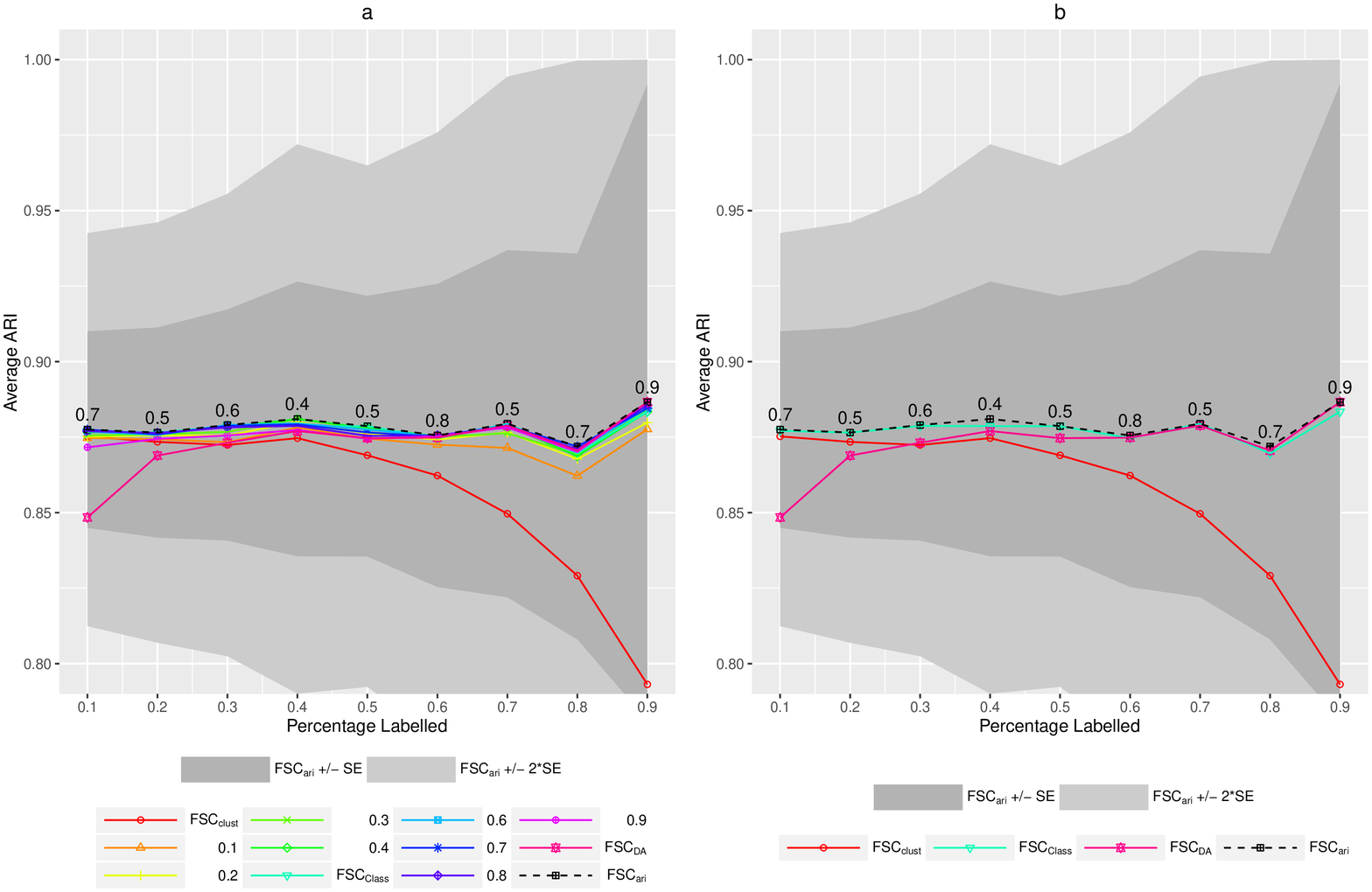}
\caption[$t$-Distribution Results for $\Delta=4$]{For $\Delta=4$: a) FSC\tsub{$\alpha$} and FSC\tsub{\tiny ARI} for $\alpha\in\alphset$, b) FSC\tsub{\tiny clust},FSC\tsub{\tiny class}, FSC\tsub{\tiny DA} and FSC\tsub{\tiny ARI}.}
\end{figure}

\subsubsection*{Estimation}
\vspace{-0.1in}
In addition to classification performance, we also consider the accuracy of the parameter estimates. Parameter estimates for FSC\tsub{\tiny ARI} from our most recent simulation are considered, for $p=$20\%, 50\%, and 80\% of points labelled and $\Delta=3$. The results (Table~\ref{tab:estimates}) show that the estimates are very close to the actual values in all cases. We note that there is a lot of variability in the estimate for $\nu_2$ --- this is to be expected because the second component is approximately Gaussian.
\begin{sidewaystable}[!h]
\centering
\caption{Average parameter estimates for $\Delta=3$ for 20\%, 50\% and 80\% of points labelled with component wise standard deviations in brackets}
\begin{tabular}{ccccccc}
\hline
$\nu_1$ (sd)& $\bmu_1$ (sd) & $\bSig_1$ (sd) &&$\nu_2$ (sd)& $\bmu_2$ (sd) & $\bSig_2$ (sd) \\
\cline{1-3}\cline{5-7}
&&&20\%&&&\\
&&&($\alpha=0.6$)&&&\\
\hline
\shortstack{$3.21$ \\$(0.766)$}& \shortstack{$\left[\begin{array}{c}-0.00698\\-0.00352\end{array}\right]$
\\ 
$\left(\left[\begin{array}{c}0.100\\0.100\end{array}\right]\right)$} &
\shortstack{$
\left[
\begin{tabular}{ll}
1.01 & 0.703\\
0.703 & 1.01\\
\end{tabular}
\right]
$
\\
$
\left(
\left[
\begin{tabular}{ll}
0.200 & 0.154\\
0.154 & 0.184\\
\end{tabular}
\right]
\right)
$
}
&&
\shortstack{$63.2$\\$ (57.0)$}& \shortstack{$\left[\begin{array}{c}0.00535\\2.99\end{array}\right]$
\\ 
$\left(\left[\begin{array}{c}0.0772\\0.0845\end{array}\right]\right)$} &
\shortstack{$
\left[
\begin{tabular}{ll}
0.988 & -0.00720\\
-0.00720 & 0.978\\
\end{tabular}
\right]
$
\\
$
\left(
\left[
\begin{tabular}{ll}
0.133 & 0.0881\\
0.0881 & 0.138\\
\end{tabular}
\right]
\right)
$
}\\
\hline
\multicolumn{7}{c}{50\%}\\
\multicolumn{7}{c}{($\alpha=0.6$)}\\
\hline
\shortstack{$3.19$ \\$(0.742)$}& \shortstack{$\left[\begin{array}{c}0.00186\\0.00476\end{array}\right]$
\\ 
$\left(\left[\begin{array}{c}0.0956\\0.0913\end{array}\right]\right)$} &
\shortstack{$
\left[
\begin{tabular}{ll}
1.03 & 0.716\\
0.716 & 1.03\\
\end{tabular}
\right]
$
\\
$
\left(
\left[
\begin{tabular}{ll}
0.195 & 0.143\\
0.143 & 0.170\\
\end{tabular}
\right]
\right)
$
}
&&
\shortstack{$67.3$\\$ (57.4)$}& \shortstack{$\left[\begin{array}{c}-0.00270\\3.00\end{array}\right]$
\\ 
$\left(\left[\begin{array}{c}0.0760\\0.0799\end{array}\right]\right)$} &
\shortstack{$
\left[
\begin{tabular}{ll}
0.990 & 0.000940\\
0.000940 & 0.980\\
\end{tabular}
\right]
$
\\
$
\left(
\left[
\begin{tabular}{ll}
0.127 & 0.0811\\
0.0811 & 0.140\\
\end{tabular}
\right]
\right)
$
}\\
\hline
\multicolumn{7}{c}{80\%}\\
\multicolumn{7}{c}{($\alpha=0.4$)}\\
\hline
\shortstack{$3.20$ \\$(0.716)$}& \shortstack{$\left[\begin{array}{c}-0.00242\\0.00122\end{array}\right]$
\\ 
$\left(\left[\begin{array}{c}0.0951\\0.0906\end{array}\right]\right)$} &
\shortstack{$
\left[
\begin{tabular}{ll}
1.02 & 0.712\\
0.712 & 1.02\\
\end{tabular}
\right]
$
\\
$
\left(
\left[
\begin{tabular}{ll}
0.194& 0.149\\
0.149 & 0.170\\
\end{tabular}
\right]
\right)
$
}
&&
\shortstack{$67.1$\\$ (52.7)$}& \shortstack{$\left[\begin{array}{c}-0.00254\\3.00\end{array}\right]$
\\ 
$\left(\left[\begin{array}{c}0.0730\\0.0737\end{array}\right]\right)$} &
\shortstack{$
\left[
\begin{tabular}{ll}
0.995 & -0.00215\\
-0.00215 & 0.978\\
\end{tabular}
\right]
$
\\
$
\left(
\left[
\begin{tabular}{ll}
0.124 & 0.0804\\
0.0804 & 0.120\\
\end{tabular}
\right]
\right)
$
}\\
\hline
\end{tabular}
\label{tab:estimates}
\end{sidewaystable}

\subsection{Simulation with Three Groups}
\begin{figure}[!h]
\centering
\includegraphics[width=0.5\textwidth,height=0.41\textwidth]{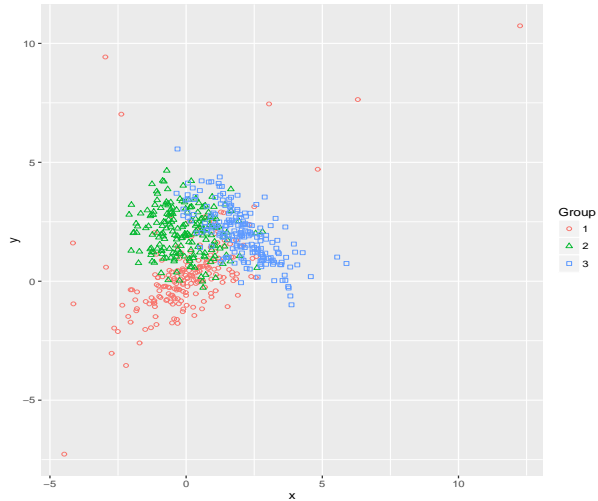}
\caption{Typical dataset for simulation with three groups.}
\label{fig:Sim3}
\end{figure}

\begin{figure}
\centering
\includegraphics[width=0.62\textwidth]{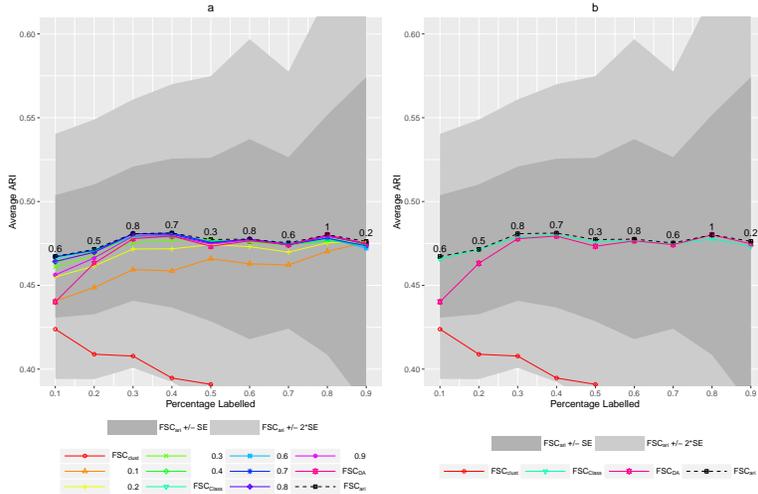}
\caption[Results for Simulation with Three Groups]{For the simulation with three groups: a) FSC\tsub{$\alpha$} and FSC\tsub{\tiny ARI} for $\alpha\in\alphset$, b) FSC\tsub{\tiny clust},FSC\tsub{\tiny class}, FSC\tsub{\tiny DA} and FSC\tsub{\tiny ARI}.}
\label{fig:ARI3}
\end{figure}

Finally, we perform a simulation with three groups. We follow the same procedure as the simulations previously discussed, this time with 100 observations in each group for a total of 300 observations for each of the 100 datasets, once again all from bivariate $t$-mixtures. The first two groups are simulated from exactly the same distributions as the previous simulations with $\Delta=2$. For the third group, we took $\vecmu_3=(2,2)$, $\nu_3=10$ and 
$$
{\mathbf \Sigma}_3=
\left[
\begin{array}{cc}
1&-0.7\\
-0.7&1\\
\end{array}
\right].
$$
A typical dataset is shown in \figurename~\ref{fig:Sim3}, where the three groups are moderately well separated but there this is still some overlap. In \figurename~\ref{fig:ARI3}, we show line plots, as before, and see that one of the three species is selected in only two of the nine cases. 
 \clearpage

\subsection{Application to Datasets}
We now look at a few datasets and compare the performance of FSC using a $t$-mixture and FSC with a Gaussian mixture. We took 100 random splits for each dataset for each percentage of labelled data, $p\in\{10,20,\ldots, 80,90\}$. We used the same criterion as in the simulations, i.e., the ARI, to choose the optimal weight. As with the simulations we use a completely unconstrained model for both the covariance structure and, in the case of the $t$-mixtures, the degrees of freedom. For completeness, we note that we are not necessarily able to perform a discriminant analysis when the percent labelled is low or a cluster analysis when the percent labelled is high.

\subsubsection*{Iris Data}
\begin{figure}[!htb]
\centering
\includegraphics[width=0.9\textwidth]{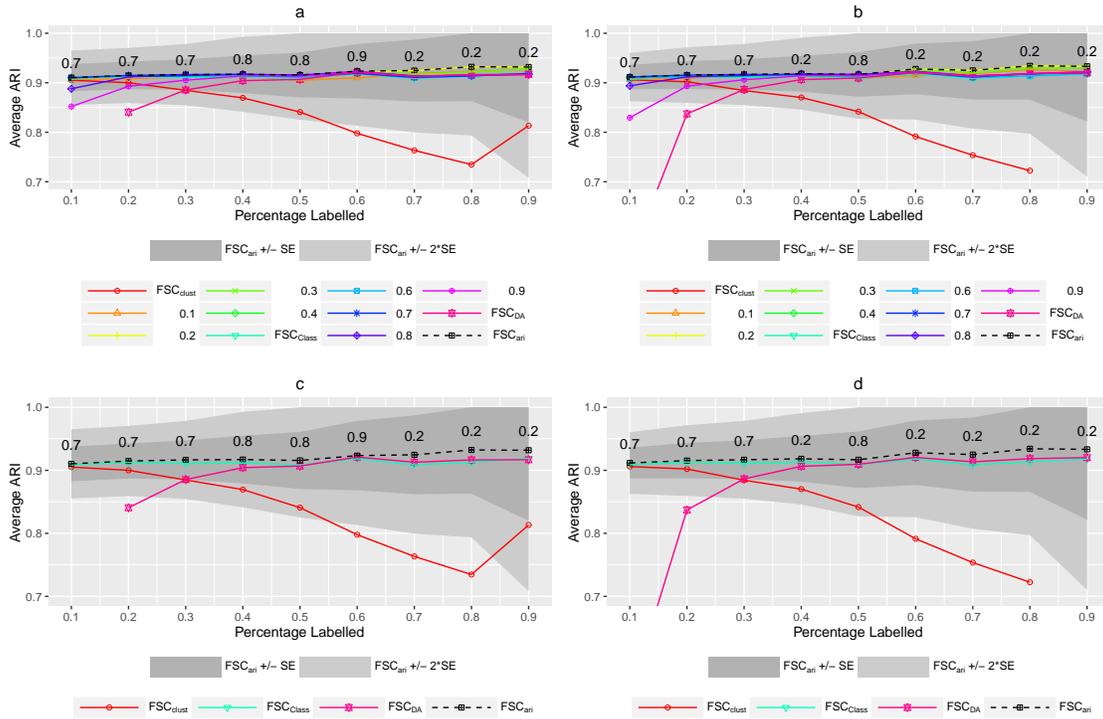}
\caption[Iris Data FSC Results]{FSC\tsub{$\alpha$} for $\alpha\in\alphset$ and FSC\tsub{\tiny ARI} for the iris data for: a) the $t$-mixture and b) for the Gaussian mixture. FSC\tsub{\tiny clust},FSC\tsub{\tiny class}, FSC\tsub{\tiny DA} and FSC\tsub{\tiny ARI} for c) the $t$-mixture, and d) the Gaussian mixture.}
\label{fig:iris}
\end{figure}

\vspace{-0.1in}
The Anderson Iris data contains four different attributes of three different species of iris and is available in the {\sf R} package {\tt datasets}. The measurements (in centimetres) are the sepal length and width, and the petal length and width. The results are depicted in \figurename~\ref{fig:iris}. On the left hand side, we show the results for the $t$-mixture, and on the right hand side we show the results for the Gaussian mixture. Comparing these two plots, we see that the overall classification performance is similar between the $t$-mixture and the Gaussian mixture. Moreover, except at $p=60$\%, the weights chosen for both the $t$ and Gaussian mixtures are very similar if not exactly the same.

\subsubsection*{Crabs Data}
\begin{figure}[!htb]
\centering
\includegraphics[width=0.9\textwidth]{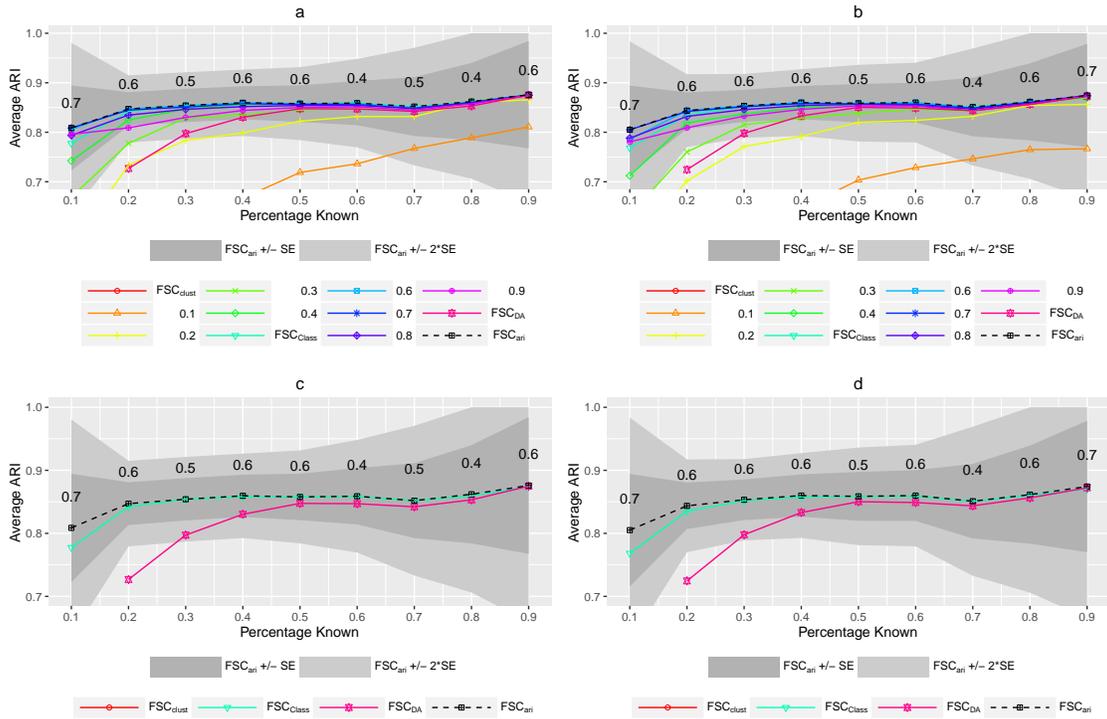}
\caption[Crab Data FSC Results]{FSC\tsub{$\alpha$} for $\alpha\in\alphset$ and FSC\tsub{\tiny ARI} for the crabs data for: a) the $t$-mixture and b) for the Gaussian mixture. FSC\tsub{\tiny clust},FSC\tsub{\tiny class}, FSC\tsub{\tiny DA} and FSC\tsub{\tiny ARI} for c) the $t$-mixture, and d) the Gaussian mixture.}
\label{fig:crab}
\end{figure}

\vspace{-0.1in}
The crabs dataset consists of 5 measurements on four different types of rock crabs (two species, male and female in each species) and are available in the {\sf R} package {\tt MASS} \citep{MASS}. These measurements are the frontal lobe size, carapace length and width, and the rear length and width. The results (\figurename~\ref{fig:crab}) show that, as for the iris data, the classification performance for the $t$ and Gaussian mixtures are similar. Moreover, the weights chosen are very similar. It is interesting to note that almost all the weights are around 0.5

\subsubsection*{Wine Data}
\begin{figure}[!htb]
\centering
\includegraphics[width=0.9\textwidth]{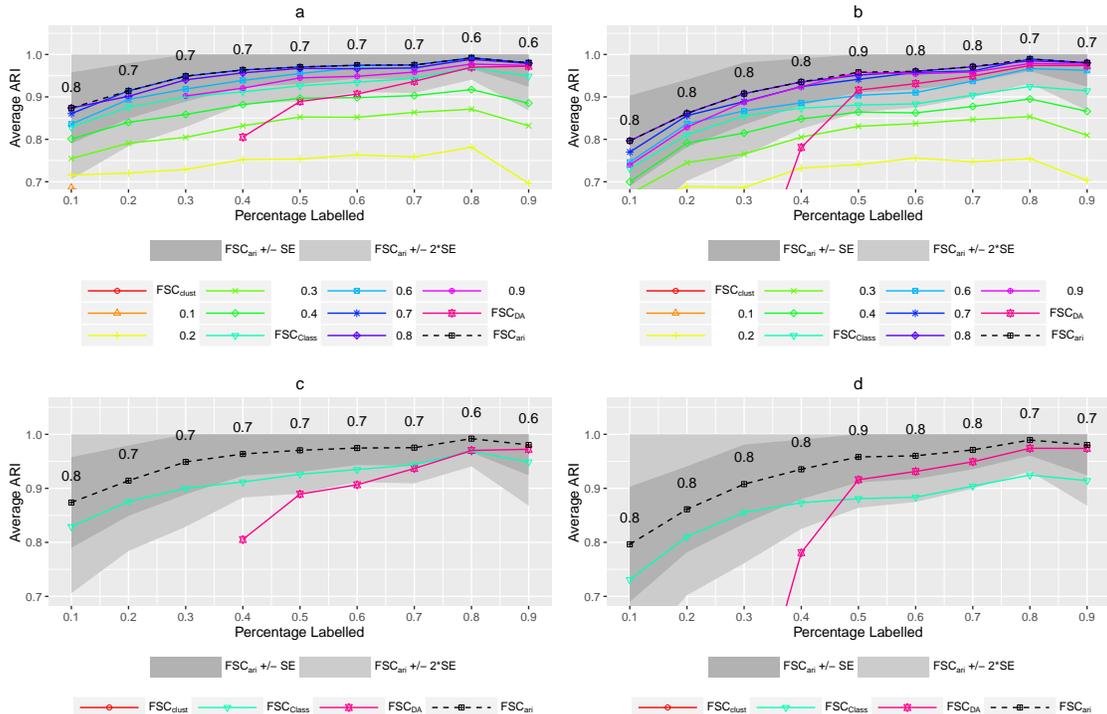}
\caption[Wine Data FSC Results]{FSC\tsub{$\alpha$} for $\alpha\in\alphset$ and FSC\tsub{\tiny ARI} for the wine data for: a) the $t$-mixture and b) for the Gaussian mixture. FSC\tsub{\tiny clust},FSC\tsub{\tiny class}, FSC\tsub{\tiny DA} and FSC\tsub{\tiny ARI} for c) the $t$-mixture, and d) the Gaussian mixture.}
\label{fig:wine}
\end{figure}

\vspace{-0.1in}
The wine dataset from the {\sf R} package {\tt gclus} \citep{hurley04} considers 13 characteristics of three different classes of wine. One interesting aspect of the results (\figurename~\ref{fig:wine}) is that, until one gets to the higher proportions of labelled data, the $t$-mixture performs slightly better than the Gaussian mixture. Another thing to note is that, similar to the crabs data, the cluster analysis does not perform well in comparison to the other values of $\alpha$. Finally, the chosen weights for the $t$- and Gaussian mixtures are fairly similar and tend to choose larger weights for the labelled observations at all proportions. 

\subsubsection*{Bankruptcy Data}
\begin{figure}[!htb]
\centering
\includegraphics[width=0.9\textwidth]{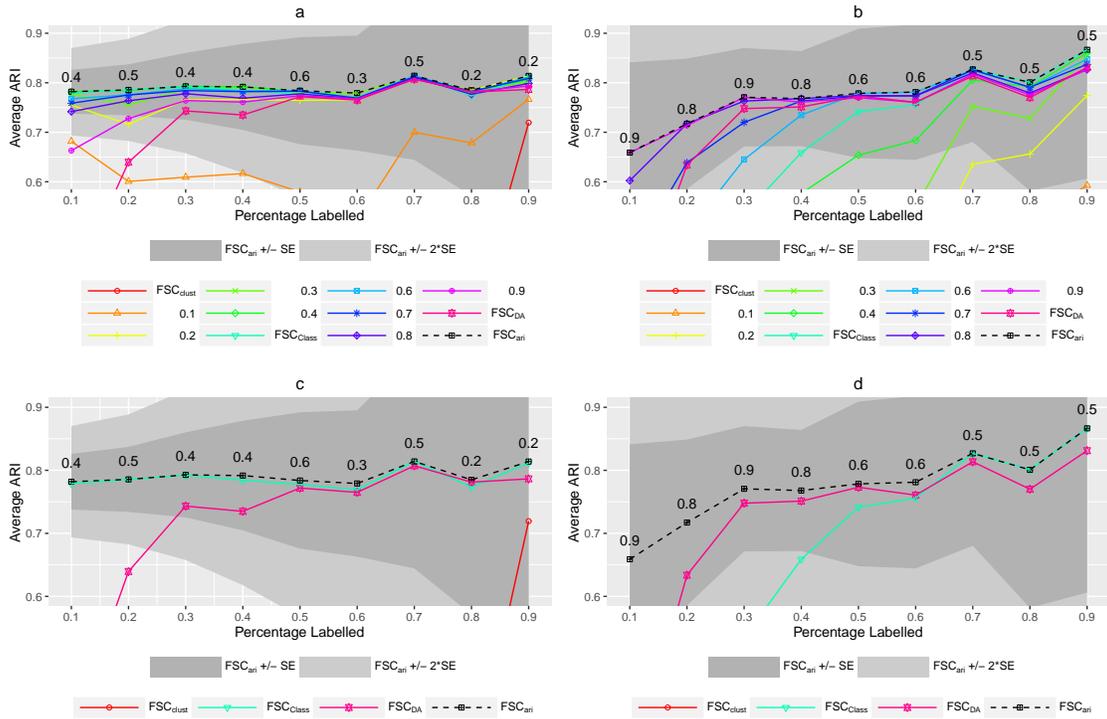}
\caption[Bankruptcy Data FSC Results]{FSC\tsub{$\alpha$} for $\alpha\in\alphset$ and FSC\tsub{\tiny ARI}  for the bankruptcy data for: a) the $t$-mixture and b) for the Gaussian mixture. FSC\tsub{\tiny clust},FSC\tsub{\tiny class}, FSC\tsub{\tiny DA} and FSC\tsub{\tiny ARI} for c) the $t$-mixture, and d) the Gaussian mixture.}
\label{fig:Bank}
\end{figure}

\vspace{-0.1in}
The bankruptcy data, found in the {\sf R} package {\tt MixGHD} \citep{tortora15c}, consider the financial situation of 66 American firms: each firm was labelled as either bankrupt or financially sound. The results (\figurename~\ref{fig:Bank}) show a greater difference between the $t$- and Gaussian mixtures when compared to the other datasets we have looked at. First, note the chosen weights. The weights chosen using a $t$-mixture are very different than those chosen when using the Gaussian mixture. The second item to note is that, similar to the wine data, the $t$-mixture gives better classification performance at lower percentages of labelled points. Finally, we note the difference in variability. For the Gaussian mixture, at lower percentages, we see a lot more variability in the error bars than for the $t$-mixture. Also, in general, there is more variability between the different weights for the Gaussian mixture. This could suggest that the selection of the weight should be treated a bit more carefully for the Gaussian mixture in this case, as the selection of a non-optimal weight can result in decreased classification performance. This is especially true, once again, at lower percentages of labelled points.

\subsection{Weight Selection Criteria for Parsimonious Models}\label{sec:weightex}
\begin{figure}[!ht]
\centering
\includegraphics[width=0.45\textwidth]{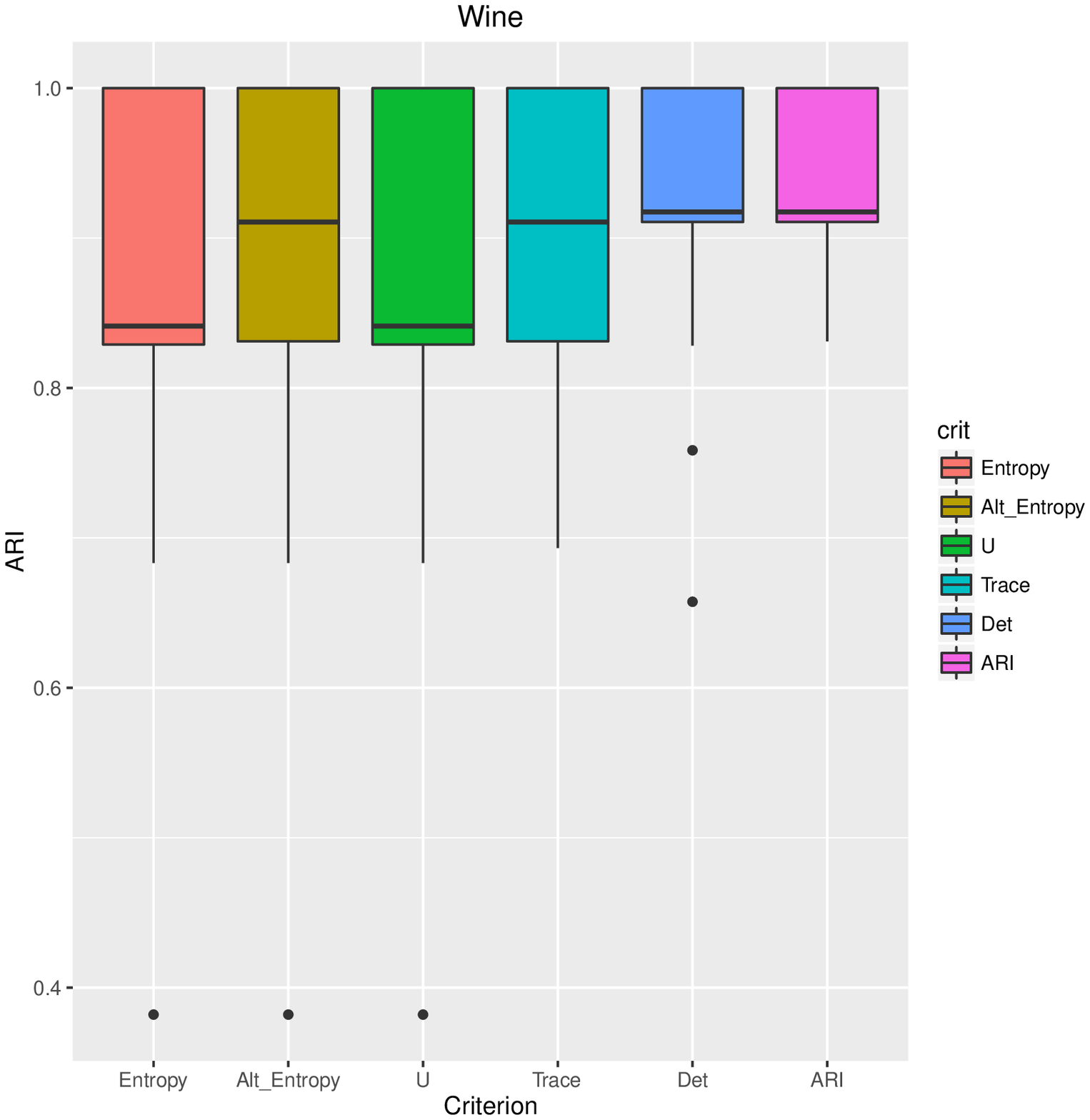}\includegraphics[width=0.45\textwidth]{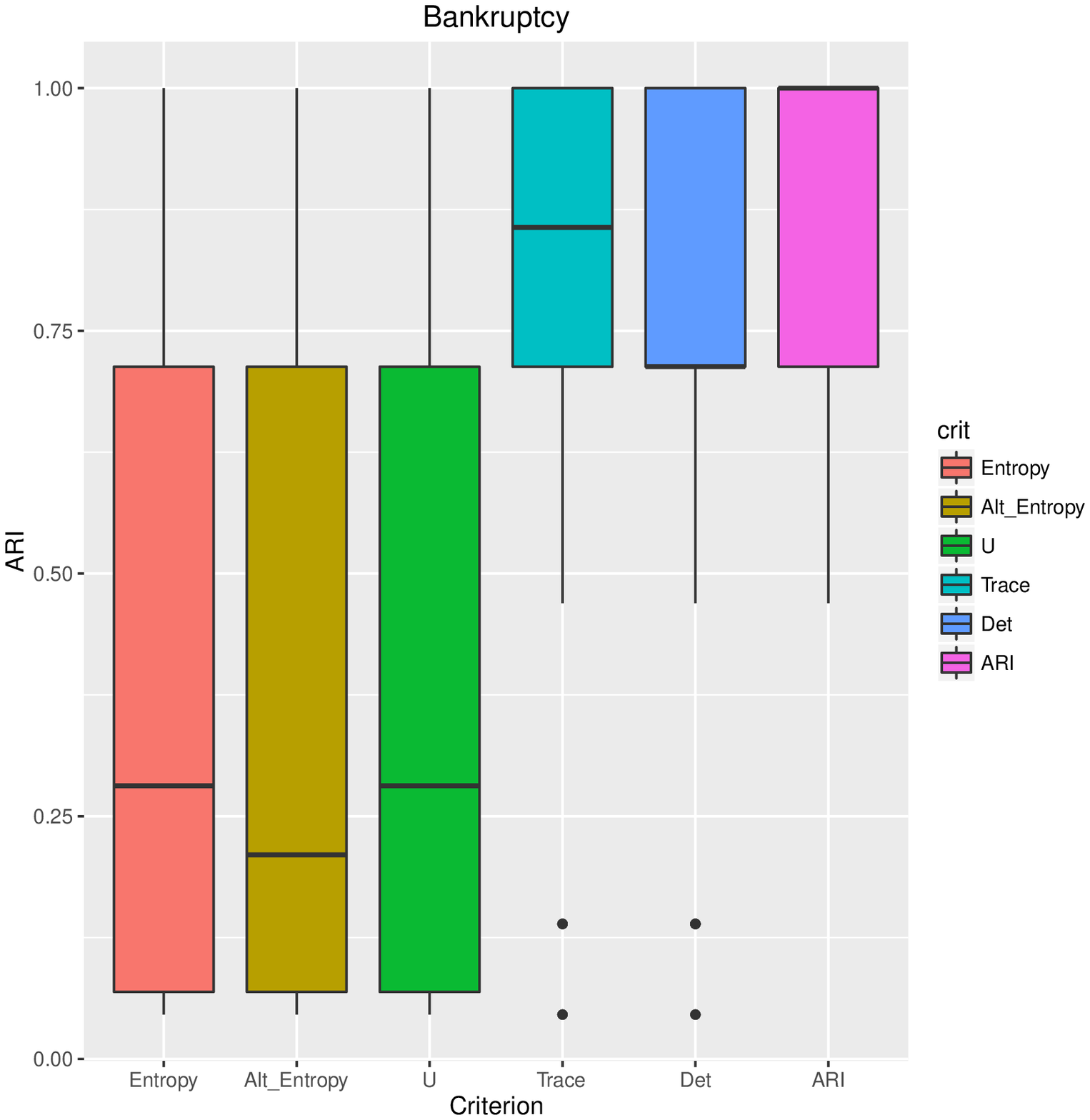}
\includegraphics[width=0.45\textwidth]{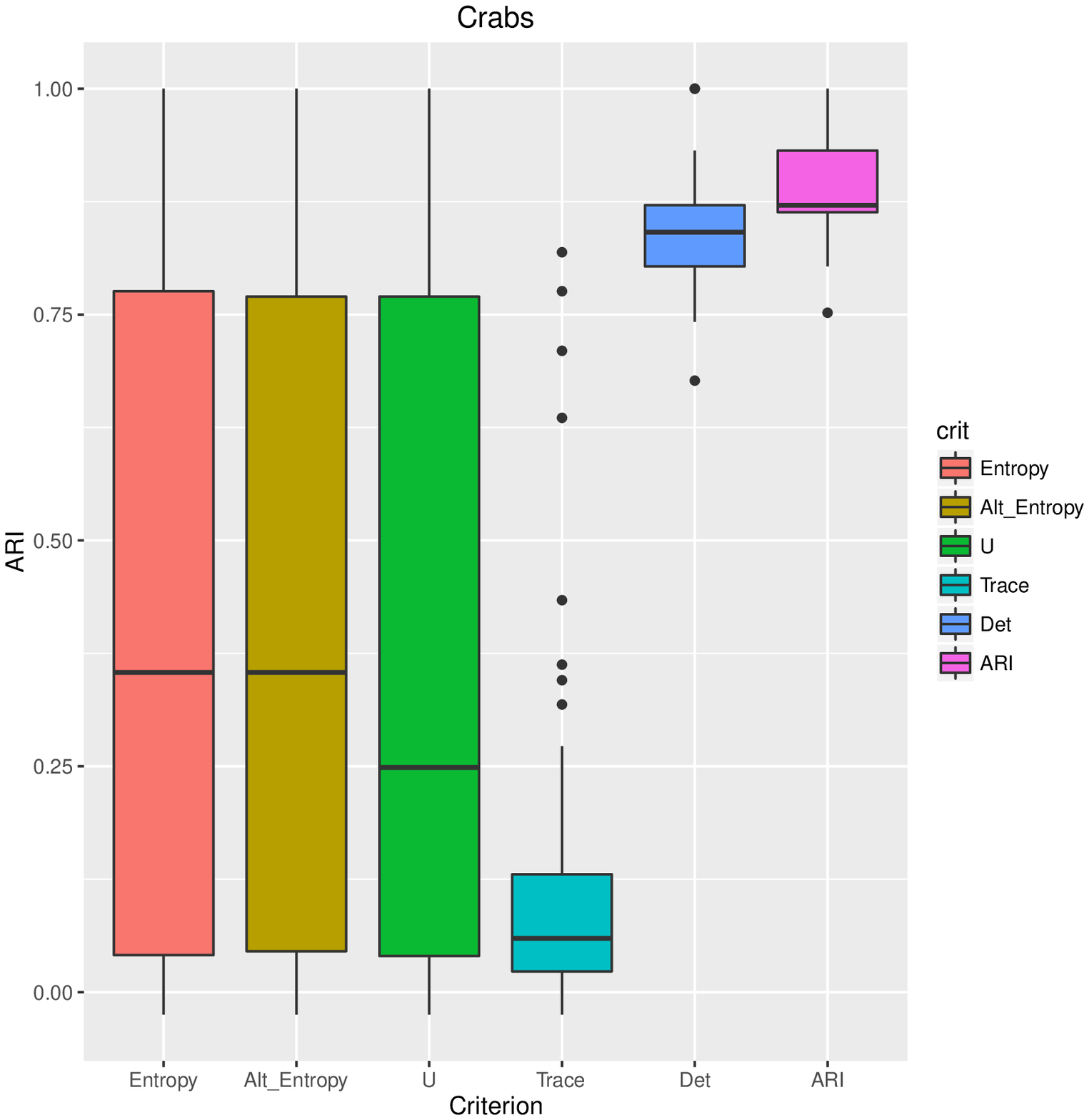}
\includegraphics[width=0.45\textwidth]{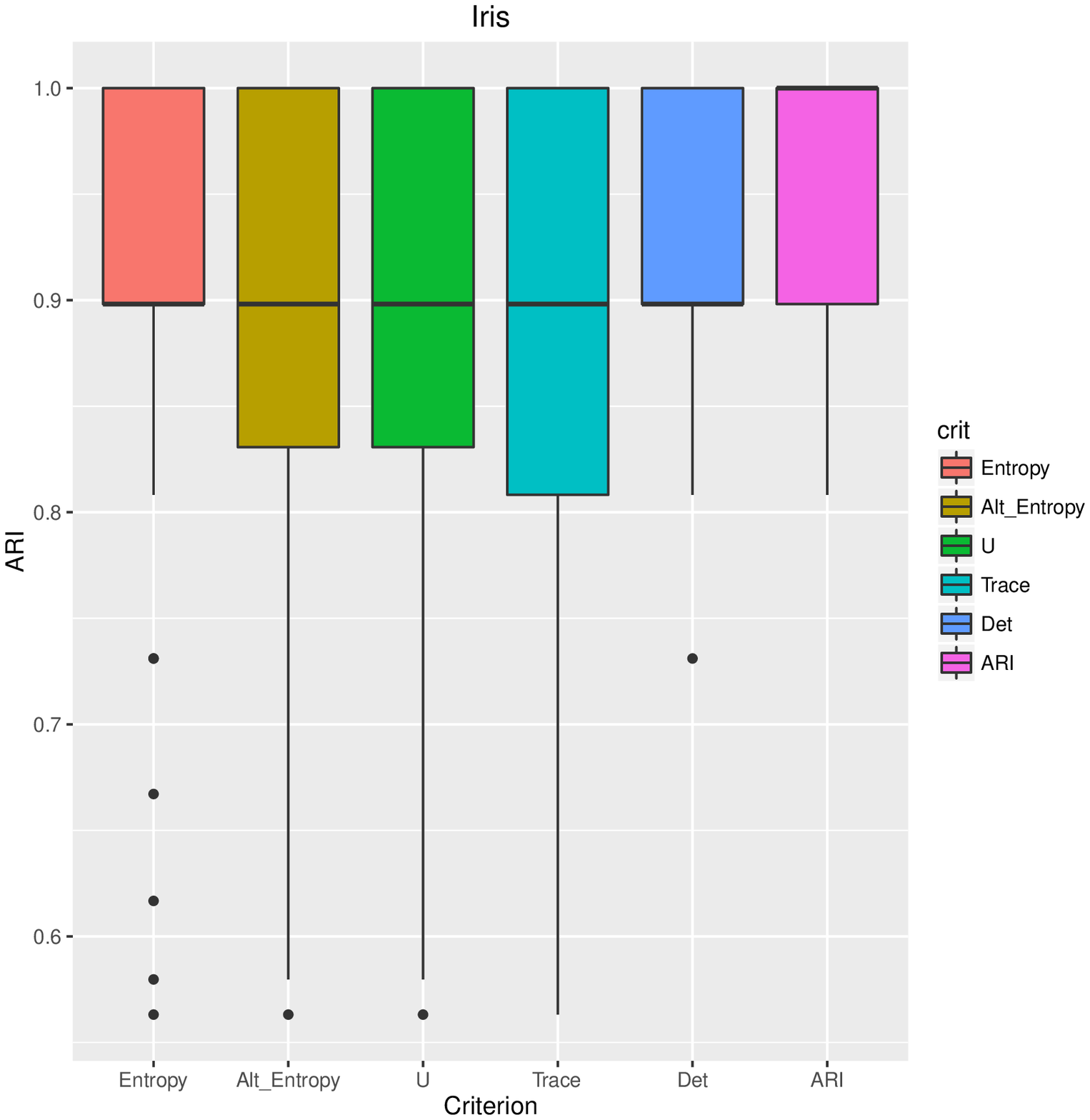}
\caption[ARI Distributions for Weight Selection Criteria]{Distribution of ARI values for each of the criteria as well as the distribution of the highest ARI for the four datasets. The BIC was used to choose the model.}
\label{fig:criterion}
\end{figure}

In Section~\ref{Method2}, five different weight selection criteria are discussed. In this section, we compare the performance of these criteria by considering FSC on $t$-mixtures for the wine, bankruptcy, crabs and iris datasets. We take 50 different splits for each dataset, with 80\% of data labelled and use a mixture of multivariate $t$-distributions. We take the same candidate weights as before (see Section~\ref{sec:sims}). For each candidate weight, we choose the model --- i.e., the value of $G$ and the covariance structure (Table~\ref{tab:teigen}, Appendix~\ref{app:tEIGEN}) --- using the BIC, and then calculate each of weight selection criteria mentioned earlier. We then choose the optimal weight, based on each of the selection criteria, and calculate the ARI. Also, we consider the highest ARI of all the weights after choosing the model to evaluate the overall performance of each of the criteria. In \figurename~\ref{fig:criterion}, we show box plots of the resulting ARI values using each of the criteria, as well as the box plot for the distribution of the highest ARI. 

The distributions of the ARI values for the three classification-based criteria show that the resulting ARI from the chosen weight is generally much lower than if we were to use the highest ARI. Moreover, the variability is generally much higher and especially so for the bankruptcy and crabs data. 
On the other hand, $\text{tr}(\mathbf{W})$ performs well in comparison to the three classification-based criteria for the wine and bankruptcy data. Furthermore, in the case of the bankruptcy data, it performs the best of all five criteria, when comparing the medians, and has a distribution closest to that of the highest ARI. However, in the case of the crabs data, it performs very poorly, and has the worst performance of the five criteria. For the iris data, the performance is similar to the alternative entropy and $U$ criteria. 
Finally, we see that $\text{det}(\mathbf{W})$ performs well for all of the datasets. In the case of the wine data, except for a couple of outliers, the distribution is very similar to that for the highest ARI --- this is quite remarkable when one considers that the ARI assumes knowledge of the true labels. Furthermore, $\text{det}(\mathbf{W})$ performs the best of all of the proposed criteria in all of the datasets except for the bankruptcy data. In this case, $\text{tr}(\mathbf{W})$ performs better, but the inter-quartile ranges are very similar. Therefore, we propose $\text{det}(\mathbf{W})$ as a criterion to select the weight $\alpha$ in FSC.

\subsubsection*{The Determinant as a Model Selection Criterion}
\begin{figure}[!htb]
\centering
\includegraphics[width=0.495\textwidth]{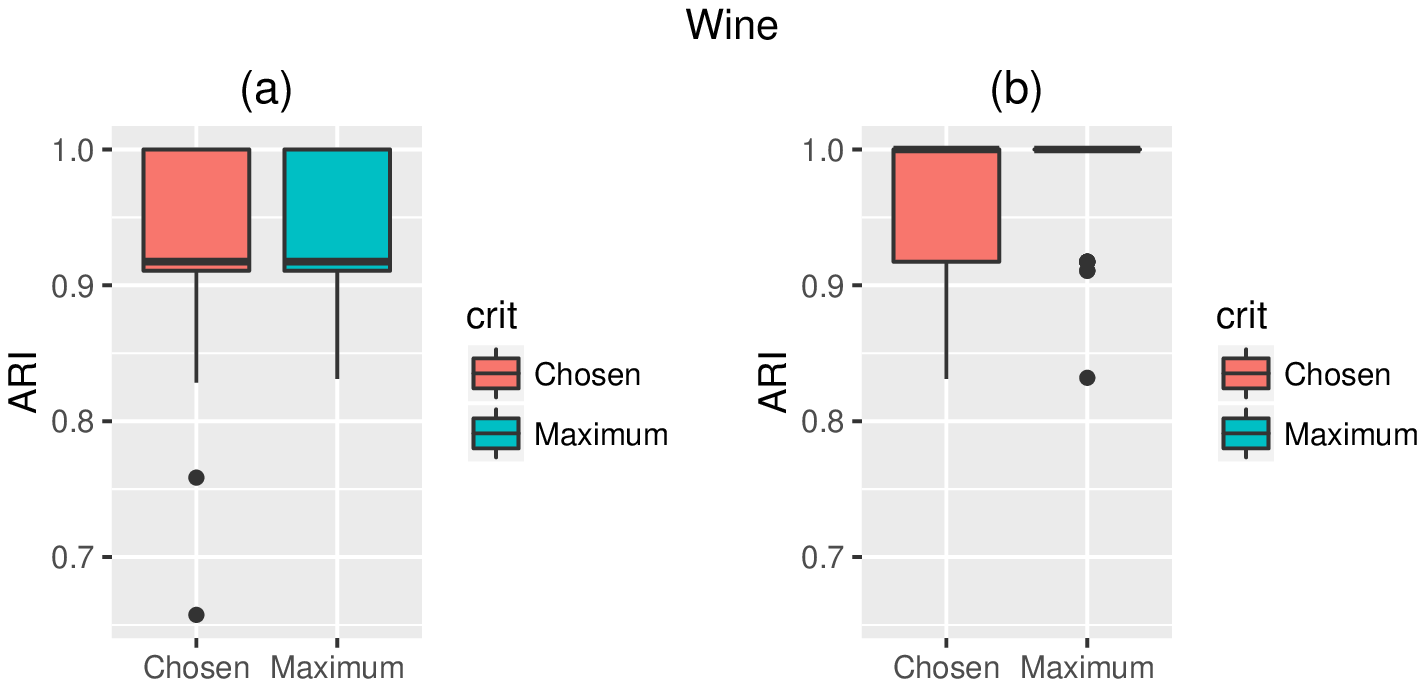}
\includegraphics[width=0.495\textwidth]{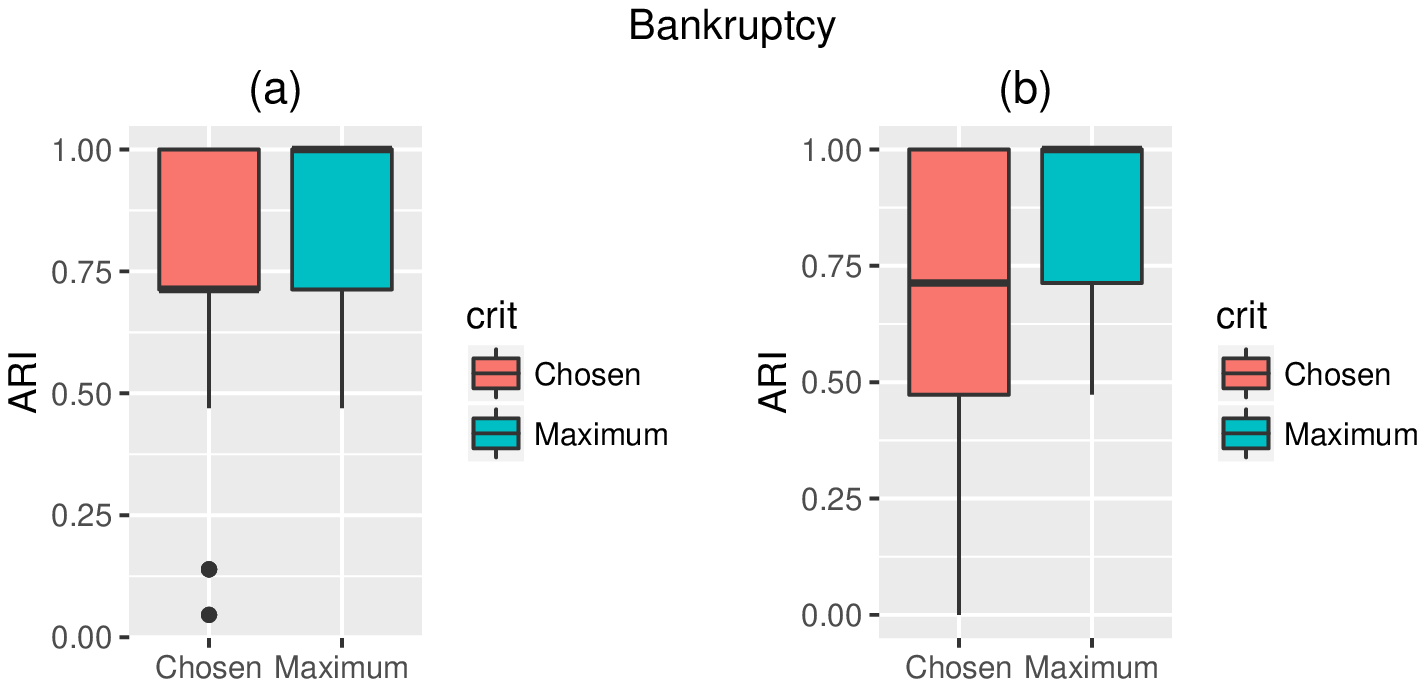}\\
\includegraphics[width=0.495\textwidth]{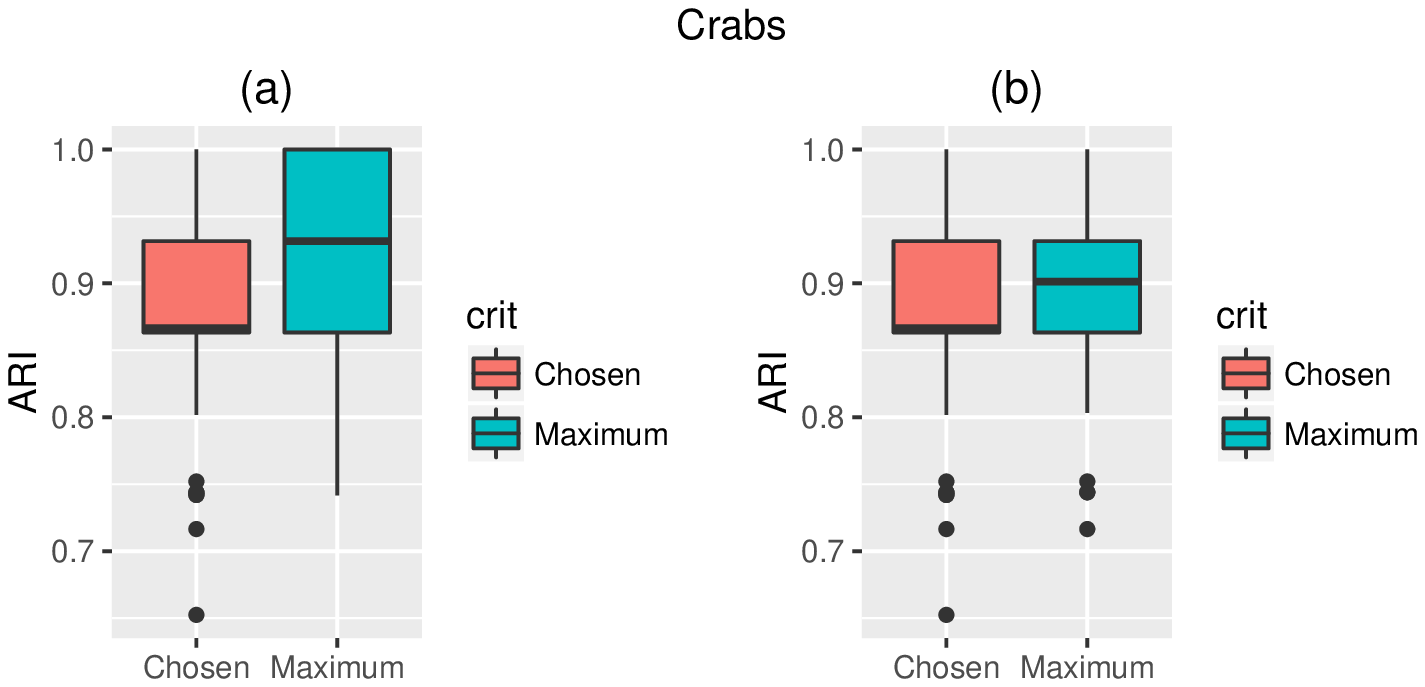}
\includegraphics[width=0.495\textwidth]{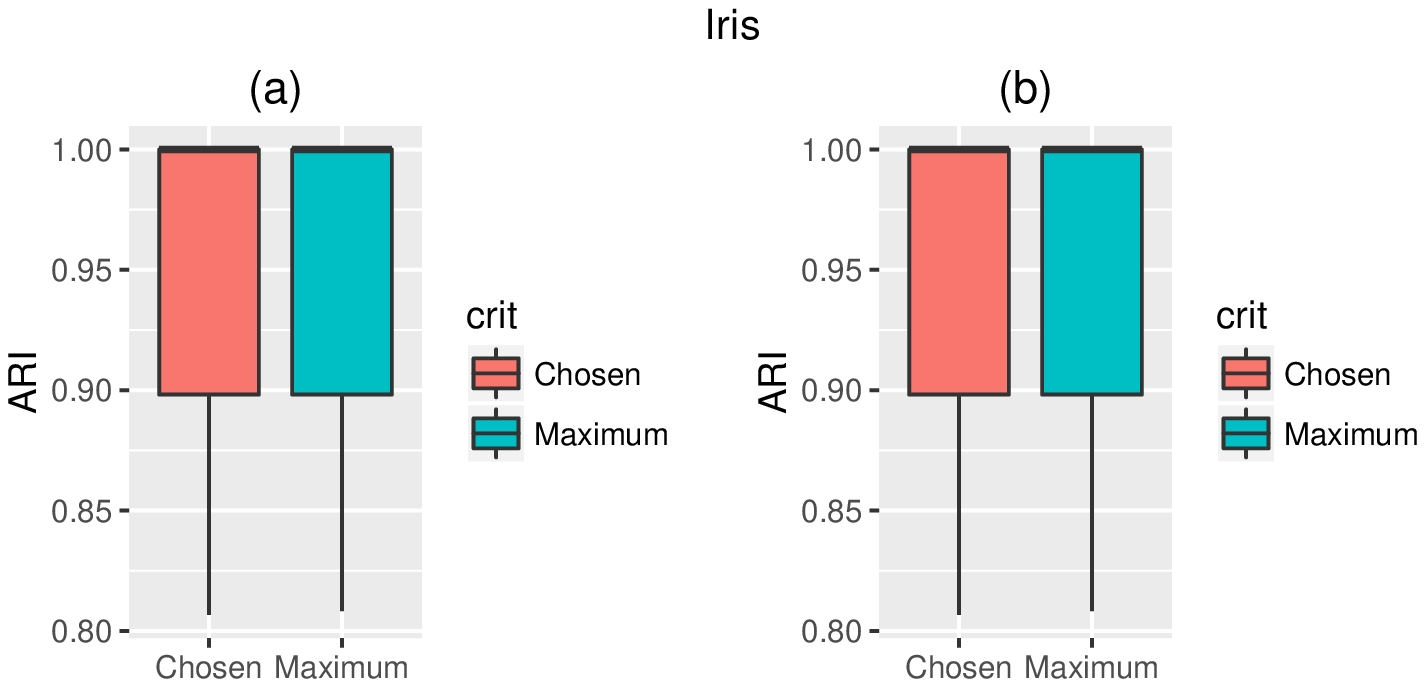}
\caption[ARI Distributions for Procedures 1 and 2]{Distribution of ARI values for (a) the first procedure and (b) the second procedure for each of the four datasets.}
\label{fig:detcrit}
\end{figure}

\vspace{-0.1in}
We have already seen that $\text{det}(\mathbf{W})$ appears to be an effective selection criterion for the weight in FSC. Now, we consider the possibility of using this criterion for model selection in general. To further explore this idea, we once again consider the four datasets and perform 50 random splits with 80\% of the data points having known labels. This time, we consider two different procedures. In the first procedure, we proceed as before and choose the model based on the BIC, and then the weight using $\text{det}(\mathbf{W})$. In the second, we choose the model based on $\text{det}(\mathbf{W})$ and then the weight also based on $\text{det}(\mathbf{W})$. We once again take the ARI values after choosing the model and the weight using one of these two procedures, and we take the maximum ARI value amongst all of the weights.
In \figurename~\ref{fig:detcrit}, we show box plots of the distributions of the ARI values. In (a) we show the results for the first procedure and, in (b), we show the results for the second procedure. 

There are a few interesting items to note. First, for the wine dataset, we see that when using $\text{det}(\mathbf{W})$ to choose the model, the distribution of the maximum ARI has a lot less variability. Also, these maximum ARI values are generally larger after using $\text{det}(\mathbf{W})$ to choose the model. One final note on the wine dataset is that the median ARI values using procedure~2 is higher than those from procedure~1.
For the bankruptcy data, we see that the distribution of the maximum ARI is the same regardless of using the BIC or $\text{det}(\mathbf{W})$ to choose the model. However, after choosing the weight, we see that the distribution of the ARI values for procedure~2 shows more variability than procedure~1. 
In the case of the crabs data, we see that the distribution of the ARI for the selected model and weight are approximately the same for both procedures; however, the maximum ARI is generally better when using the BIC to choose the model.
Finally, for the Iris data, all of the distributions are very similar.
The results are inconclusive in that neither procedure outperformed the other; however, the fact that the BIC did not outperform $\text{det}(\mathbf{W})$ for model selection is remarkable. In fact, the possibility of using $\text{det}(\mathbf{W})$ for model selection in model-based clustering, as alternative to the BIC, is worthy of further consideration.

\subsection{Justification for a Cluster Analysis}
\begin{figure}[!htb]
\centering
\includegraphics[width=1.0\textwidth]{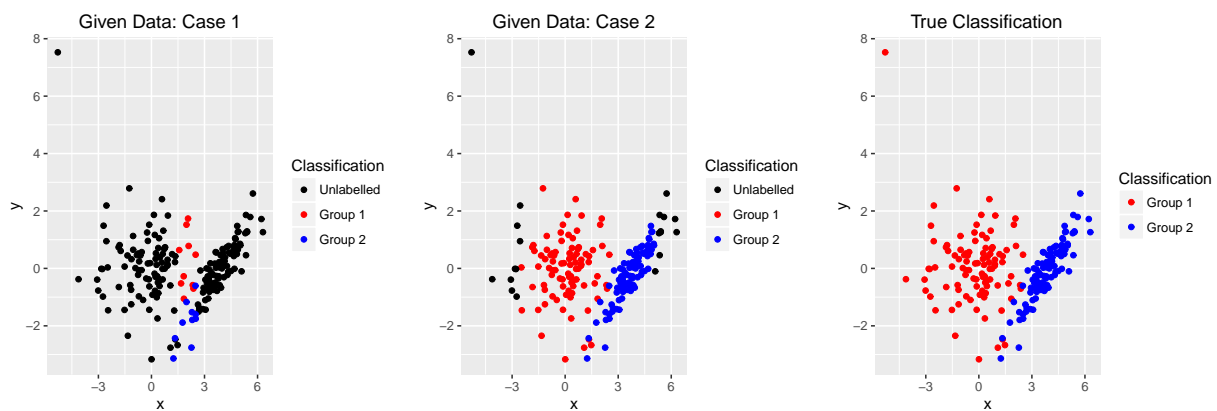}
\caption[Datasets for Cluster Analysis Justification]{Two different possible datasets with different organizations of labelled points with the true classification.}
\label{fig:ClustCase}
\end{figure}

If some of the points are labelled, it may not be immediately clear as to why a cluster analysis should even be considered. However, there are situations in which performing a cluster analysis is just as good, if not better, than putting more weight on the labelled observations. In \figurename~\ref{fig:ClustCase}, we show two different situations where this would be the case. 
In Table \ref{tab:aricase}, we look at the ARI and $\text{det}(\mathbf{W})$ for each of the weights for the two different cases. In the first case, only 10\% of the points are labelled, and all labelled points are around the intersection of the two clusters. In this case, we see from the ARI and determinant values that we would only want give very little weight, or no weight, to the labelled observations.
\begin{table}[!htb]
\centering
\caption{ARI and determinant values for each candidate weight for both of the cases in \figurename~\ref{fig:ClustCase}.}
\begin{tabular}{lrrrrrr}
\hline
 && \multicolumn{2}{c}{First Case} && \multicolumn{2}{c}{Second Case}\\
\cline{3-4}\cline{6-7}
Weight&& ARI & Det. && ARI & Det.\\
\hline
0 && 0.9341 & 81006 && 1 & 82849 \\ 
  0.1 && 0.9341 & 81006 && 1 & 82849 \\ 
  0.2 && 0.9341 & 81006 && 1 & 82849 \\ 
  0.3 && 0.9126 & 81984 && 1 & 82849 \\ 
  0.4 && 0.9126 & 81984 && 1 & 82849 \\ 
  0.5 && 0.9126 & 81984 && 1 & 82849 \\ 
  0.6 && 0.8914 & 84250 && 1 & 82849 \\ 
  0.7 && 0.8914 & 84250 && 1 & 82849 \\ 
  0.8 && 0.8914 & 84250 && 1 & 82849 \\ 
  0.9 && 0.0075 & 178858 && 1 & 82849 \\ 
  1 && $-0.0016$ & 187192 && 1 & 82849 \\ 
 \hline
\end{tabular}
\label{tab:aricase}
\end{table}

In the first case, we see that a cluster analysis is actually better than using higher weights, and just as good as using smaller weights. In the second case, 90\% of the points are labelled, and the unlabelled points lie on the outside of the two clusters. From the ARI and $\text{det}(\mathbf{W})$ values (Table~\ref{tab:aricase}), it is clear that all weights give perfect classification, including a cluster analysis, and thus a cluster analysis would perform just as well as the other weights in this case.

\section{Conclusions and Future Work}\label{sec:disc}
The major contribution of this paper is to encourage the use of $\text{det}(\mathbf{W})$ as a weight selection criterion in FSC. Although based on old ideas, and ideas that have not been fashionable for some time, this criteria is shown to outperform alternatives such as the near-ubiquitous BIC for weight selection. Furthermore, it performs comparably to the BIC in the model selection stage. As a secondary contribution, the FSC approach is shown to be mathematically tractable and effective for mixtures of multivariate $t$-distributions. For example, in our simulations, the selected weight very rarely corresponded to one of the three traditional species of classification. Furthermore, in our real data analyses, the use of a mixture of multivariate $t$-distributions was shown to either perform as well as or, in the case of the wine and bankruptcy datasets, better than the mixture of multivariate Gaussian distributions. This is likely due, at least in part, to the $t$-distribution being more robust to outliers than the Gaussian distribution. It is not unreasonable to expect that the FSC will also perform well with other non-Gaussian mixtures --- the reader is referred to the recent review paper of \cite{mcnicholas16b} for some discussion of non-Gaussian mixtures. 

Future work will investigate using $\text{det}(\mathbf{W})$ as an alternative to the BIC for model selection in model-based clustering and classification in general. Using the FSC approach in a wider range of situations will also be explored. For example, FSC could be applied in the area of item response theory.

\subsection*{Acknowledgements}
\vspace{-0.1in}
This work is supported by a Vanier CGS-D scholarship from the Natural Sciences and Engineering Research Council of Canada (Gallaugher) and the Canada Research Chairs program (McNicholas).

{ 

\begin{thebibliography}{}

\bibitem[\protect\citeauthoryear{Aitken}{Aitken}{1926}]{aitken26}
Aitken, A.~C. (1926).
\newblock A series formula for the roots of algebraic and transcendental equations.
\newblock {\em Proceedings of the Royal Society of Edinburgh\/}~{\em 45\/}, 14--22.

\bibitem[\protect\citeauthoryear{Andrews and McNicholas}{Andrews and
  McNicholas}{2011a}]{andrews11a}
Andrews, J.~L. and P.~D. McNicholas (2011a).
\newblock Extending mixtures of multivariate t-factor analyzers.
\newblock {\em Statistics and Computing\/}~{\em 21\/}(3), 361--373.

\bibitem[\protect\citeauthoryear{Andrews and McNicholas}{Andrews and
  McNicholas}{2011b}]{andrews11c}
Andrews, J.~L. and P.~D. McNicholas (2011b).
\newblock Mixtures of modified t-factor analyzers for model-based clustering,
  classification, and discriminant analysis.
\newblock {\em Journal of Statistical Planning and Inference\/}~{\em 141\/}(4),
  1479--1486.

\bibitem[\protect\citeauthoryear{Andrews and McNicholas}{Andrews and
  McNicholas}{2012}]{andrews12}
Andrews, J.~L. and P.~D. McNicholas (2012).
\newblock Model-based clustering, classification, and discriminant analysis via
  mixtures of multivariate $t$-distributions: The $t${EIGEN} family.
\newblock {\em Statistics and Computing\/}~{\em 22\/}(5), 1021--1029.

\bibitem[\protect\citeauthoryear{Andrews and McNicholas}{Andrews and
  McNicholas}{2014}]{andrews14}
Andrews, J.~L. and P.~D. McNicholas (2014).
\newblock Variable selection for clustering and classification.
\newblock {\em Journal of Classification\/}~{\em 31\/}(2), 136--153.

\bibitem[\protect\citeauthoryear{Andrews, Wickins, Boers, and
  McNicholas}{Andrews et~al.}{2016}]{andrews15}
Andrews, J.~L., J.~R. Wickins, N.~M. Boers, and P.~D. McNicholas (2016).
\newblock {\em {teigen}: Model-Based Clustering and Classification with the
  Multivariate t Distribution}.
\newblock R package version 2.2.0.

\bibitem[\protect\citeauthoryear{Banfield and Raftery}{Banfield and
  Raftery}{1993}]{banfield93}
Banfield, J.~D. and A.~E. Raftery (1993).
\newblock Model-based {G}aussian and non-{G}aussian clustering.
\newblock {\em Biometrics\/}~{\em 49\/}(3), 803--821.

\bibitem[\protect\citeauthoryear{Baum, Petrie, Soules, and Weiss}{Baum
  et~al.}{1970}]{baum70}
Baum, L.~E., T.~Petrie, G.~Soules, and N.~Weiss (1970).
\newblock A maximization technique occurring in the statistical analysis of
  probabilistic functions of {M}arkov chains.
\newblock {\em Annals of Mathematical Statistics\/}~{\em 41}, 164--171.

\bibitem[\protect\citeauthoryear{Bensmail, Celeux, Raftery, and
  Robert}{Bensmail et~al.}{1997}]{bensmail97}
Bensmail, H., G.~Celeux, A.~Raftery, and C.~Robert (1997).
\newblock Inference in model-based cluster analysis.
\newblock {\em Statistics and Computing\/}~{\em 7}, 1--10.

\bibitem[\protect\citeauthoryear{Biernacki, Celeux, and Govaert}{Biernacki
  et~al.}{2000}]{biernacki00}
Biernacki, C., G.~Celeux, and G.~Govaert (2000).
\newblock Assessing a mixture model for clustering with the integrated
  completed likelihood.
\newblock {\em IEEE Transactions on Pattern Analysis and Machine
  Intelligence\/}~{\em 22\/}(7), 719--725.

\bibitem[\protect\citeauthoryear{Celeux and Govaert}{Celeux and
  Govaert}{1995}]{celeux95}
Celeux, G. and G.~Govaert (1995).
\newblock Gaussian parsimonious clustering models.
\newblock {\em Pattern Recognition\/}~{\em 28\/}(5), 781--793.

\bibitem[\protect\citeauthoryear{Celeux and Soromenho}{Celeux and
  Soromenho}{1996}]{celeux96}
Celeux, G. and G.~Soromenho (1996).
\newblock An entropy criterion for assessing the number of clusters in a
  mixture model.
\newblock {\em Journal of Classification\/}~{\em 13}, 195--212.

\bibitem[\protect\citeauthoryear{Dang, Browne, and McNicholas}{Dang
  et~al.}{2015}]{dang15}
Dang, U.~J., R.~P. Browne, and P.~D. McNicholas (2015).
\newblock Mixtures of multivariate power exponential distributions.
\newblock {\em Biometrics\/}~{\em 71\/}(4), 1081--1089.

\bibitem[\protect\citeauthoryear{Dempster, Laird, and Rubin}{Dempster
  et~al.}{1977}]{dempster77}
Dempster, A.~P., N.~M. Laird, and D.~B. Rubin (1977).
\newblock Maximum likelihood from incomplete data via the {EM} algorithm.
\newblock {\em Journal of the Royal Statistical Society: Series~B\/}~{\em
  39\/}(1), 1--38.

\bibitem[\protect\citeauthoryear{Edwards and Cavalli-Sforza}{Edwards and
  Cavalli-Sforza}{1965}]{edwards65}
Edwards, A. W.~F. and L.~L. Cavalli-Sforza (1965).
\newblock A method for cluster analysis.
\newblock {\em Biometrics\/}~{\em 21}, 362--375.

\bibitem[\protect\citeauthoryear{Fraley and Raftery}{Fraley and
  Raftery}{1998}]{fraley98}
Fraley, C. and A.~E. Raftery (1998).
\newblock How many clusters? {W}hich clustering methods? {A}nswers via
  model-based cluster analysis.
\newblock {\em The Computer Journal\/}~{\em 41\/}(8), 578--588.

\bibitem[\protect\citeauthoryear{Franczak, Browne, and McNicholas}{Franczak
  et~al.}{2014}]{franczak14}
Franczak, B.~C., R.~P. Browne, and P.~D. McNicholas (2014).
\newblock Mixtures of shifted asymmetric {L}aplace distributions.
\newblock {\em IEEE Transactions on Pattern Analysis and Machine
  Intelligence\/}~{\em 36\/}(6), 1149--1157.

\bibitem[\protect\citeauthoryear{Franczak, Tortora, Browne, and McNicholas}{Franczak
  et~al.}{2015}]{franczak15}
Franczak, B.~C., C.~Tortora, R.~P. Browne, and P.~D. McNicholas (2015).
\newblock Unsupervised learning via mixtures of skewed distributions with hypercube contours.
\newblock {\em Pattern Recognition Letters\/}~{\em 58\/}(1), 69--76.

\bibitem[\protect\citeauthoryear{Friedman and Rubin}{Friedman and
  Rubin}{1967}]{friedman67}
Friedman, H.~P. and J.~Rubin (1967).
\newblock On some invariant criteria for grouping data.
\newblock {\em Journal of the American Statistical Association\/}~{\em 62},
  1159--1178.

\bibitem[\protect\citeauthoryear{Gordon}{Gordon}{1981}]{gordon81}
Gordon, A.~D. (1981).
\newblock {\em Classification}.
\newblock London: Chapman and Hall.

\bibitem[\protect\citeauthoryear{Hubert and Arabie}{Hubert and
  Arabie}{1985}]{hubert85}
Hubert, L. and P.~Arabie (1985).
\newblock Comparing partitions.
\newblock {\em Journal of Classification\/}~{\em 2\/}(1), 193--218.

\bibitem[\protect\citeauthoryear{Hurley}{Hurley}{2004}]{hurley04}
Hurley, C. (2004).
\newblock Clustering visualizations of multivariate data.
\newblock {\em Journal of Computational and Graphical Statistics\/}~{\em
  13\/}(4), 788--806.

\bibitem[\protect\citeauthoryear{Ingrassia, Minotti, Punzo, and
  Vittadini}{Ingrassia et~al.}{2015}]{ingrassia15}
Ingrassia, S., S.~C. Minotti, A.~Punzo, and G.~Vittadini (2015).
\newblock The generalized linear mixed cluster-weighted model.
\newblock {\em Journal of Classification\/}~{\em 32\/}(1), 85--113.

\bibitem[\protect\citeauthoryear{Ingrassia, Minotti, and Vittadini}{Ingrassia
  et~al.}{2012}]{ingrassia12}
Ingrassia, S., S.~C. Minotti, and G.~Vittadini (2012).
\newblock Local statistical modeling via the cluster-weighted approach with
  elliptical distributions.
\newblock {\em Journal of Classification\/}~{\em 29\/}(3), 363--401.

\bibitem[\protect\citeauthoryear{Lee and McLachlan}{Lee and
  McLachlan}{2014}]{lee14}
Lee, S. and G.~J. McLachlan (2014).
\newblock Finite mixtures of multivariate skew t-distributions: some recent and
  new results.
\newblock {\em Statistics and Computing\/}~{\em 24}, 181--202.

\bibitem[\protect\citeauthoryear{Lee and McLachlan}{Lee and
  McLachlan}{2013}]{lee13}
Lee, S.~X. and G.~J. McLachlan (2013).
\newblock On mixtures of skew normal and skew t-distributions.
\newblock {\em Advances in Data Analysis and Classification\/}~{\em 7\/}(3),
  241--266.

\bibitem[\protect\citeauthoryear{Lin}{Lin}{2010}]{lin10}
Lin, T.-I. (2010).
\newblock Robust mixture modeling using multivariate skew t distributions.
\newblock {\em Statistics and Computing\/}~{\em 20\/}(3), 343--356.

\bibitem[\protect\citeauthoryear{Lin, McLachlan, and Lee}{Lin
  et~al.}{2016}]{lin13}
Lin, T.-I., G.~J. McLachlan, and S.~X. Lee (2016).
\newblock Extending mixtures of factor models using the restricted multivariate
  skew-normal distribution.
\newblock {\em Journal of Multivariate Analysis\/}~{\em 143}, 398--413.

\bibitem[\protect\citeauthoryear{Lin, McNicholas, and Hsiu}{Lin
  et~al.}{2014}]{lin14}
Lin, T.-I., P.~D. McNicholas, and J.~H. Hsiu (2014).
\newblock Capturing patterns via parsimonious t mixture models.
\newblock {\em Statistics and Probability Letters\/}~{\em 88}, 80--87.

\bibitem[\protect\citeauthoryear{MacQueen}{MacQueen}{1967}]{macqueen67}
MacQueen, J. (1967).
\newblock Some methods for classification and analysis of multivariate
  observations.
\newblock In {\em Proceedings of the Fifth Berkeley Symposium on Mathematical
  Statistics and Probability, Volume 1: Statistics}, Berkeley, pp.\  281--297.
  University of California Press.

\bibitem[\protect\citeauthoryear{McNicholas}{McNicholas}{2016a}]{mcnicholas16a}
McNicholas, P.~D. (2016a).
\newblock {\em Mixture Model-Based Classification}.
\newblock Boca Raton: Chapman \& Hall/CRC Press.

\bibitem[\protect\citeauthoryear{McNicholas}{McNicholas}{2016b}]{mcnicholas16b}
McNicholas, P.~D. (2016b).
\newblock Model-based clustering.
\newblock {\em Journal of Classification\/}~{\em 33\/}(3), 331--373.

\bibitem[\protect\citeauthoryear{McNicholas and Murphy}{McNicholas and
  Murphy}{2008}]{mcnicholas08}
McNicholas, P.~D. and T.~B. Murphy (2008).
\newblock Parsimonious {G}aussian mixture models.
\newblock {\em Statistics and Computing\/}~{\em 18\/}(3), 285--296.

\bibitem[\protect\citeauthoryear{McNicholas, Murphy, McDaid, and
  Frost}{McNicholas et~al.}{2010}]{mcnicholas10a}
McNicholas, P.~D., T.~B. Murphy, A.~F. McDaid, and D.~Frost (2010).
\newblock Serial and parallel implementations of model-based clustering via
  parsimonious {G}aussian mixture models.
\newblock {\em Computational Statistics and Data Analysis\/}~{\em 54\/}(3),
  711--723.

\bibitem[\protect\citeauthoryear{Murray, Browne, and McNicholas}{Murray
  et~al.}{2014a}]{murray14b}
Murray, P.~M., R.~B. Browne, and P.~D. McNicholas (2014a).
\newblock Mixtures of skew-t factor analyzers.
\newblock {\em Computational Statistics and Data Analysis\/}~{\em 77},
  326--335.

\bibitem[\protect\citeauthoryear{Murray, Browne, and McNicholas}{Murray
  et~al.}{2017a}]{murray17}
Murray, P.~M., R.~B. Browne, and P.~D. McNicholas (2017a).
\newblock Hidden truncation hyperbolic distributions, finite mixtures thereof,
  and their application for clustering.
\newblock {\em Journal of Multivariate Analysis\/}~{\em 161}, 141--156.

\bibitem[\protect\citeauthoryear{Murray, Browne, and McNicholas}{Murray
  et~al.}{2017b}]{murray17b}
Murray, P.~M., R.~B. Browne, and P.~D. McNicholas (2017b).
\newblock A mixture of {SDB} skew-t factor analyzers.
\newblock {\em Econometrics and Statistics\/}~{\em 3}, 160--168.

\bibitem[\protect\citeauthoryear{Murray, McNicholas, and Browne}{Murray
  et~al.}{2014b}]{murray14a}
Murray, P.~M., P.~D. McNicholas, and R.~B. Browne (2014b).
\newblock A mixture of common skew-$t$ factor analyzers.
\newblock {\em Stat\/}~{\em 3\/}(1), 68--82.

\bibitem[\protect\citeauthoryear{Orchard and Woodbury}{Orchard and
  Woodbury}{1972}]{orchard72}
Orchard, T. and M.~A. Woodbury (1972).
\newblock A missing information principle: Theory and applications.
\newblock In L.~M. Le~Cam, J.~Neyman, and E.~L. Scott (Eds.), {\em Proceedings
  of the Sixth Berkeley Symposium on Mathematical Statistics and Probability,
  Volume~1: Theory of Statistics}, pp.\  697--715. Berkeley: University of
  California Press.

\bibitem[\protect\citeauthoryear{Peel and McLachlan}{Peel and
  McLachlan}{2000}]{peel00}
Peel, D. and G.~J. McLachlan (2000).
\newblock Robust mixture modelling using the t distribution.
\newblock {\em Statistics and Computing\/}~{\em 10\/}(4), 339--348.

\bibitem[\protect\citeauthoryear{Punzo and McNicholas}{Punzo and
  McNicholas}{2017}]{punzo17}
Punzo, A. and P.~D. McNicholas (2017).
\newblock Robust clustering in regression analysis via the contaminated
  {G}aussian cluster-weighted model.
\newblock {\em Journal of Classification\/}~{\em 34\/}(2), 249--293.

\bibitem[\protect\citeauthoryear{{R Core Team}}{{R Core Team}}{2016}]{R15}
{R Core Team} (2016).
\newblock {\em R: A Language and Environment for Statistical Computing}.
\newblock Vienna, Austria: R Foundation for Statistical Computing.

\bibitem[\protect\citeauthoryear{Schwarz}{Schwarz}{1978}]{schwarz78}
Schwarz, G. (1978).
\newblock Estimating the dimension of a model.
\newblock {\em The Annals of Statistics\/}~{\em 6\/}(2), 461--464.

\bibitem[\protect\citeauthoryear{Scott and Symons}{Scott and
  Symons}{1971}]{scott71}
Scott, A.~J. and M.~J. Symons (1971).
\newblock Clustering methods based on likelihood ratio criteria.
\newblock {\em Biometrics\/}~{\em 27}, 387--397.

\bibitem[\protect\citeauthoryear{Steane, McNicholas, and Yada}{Steane
  et~al.}{2012}]{steane12}
Steane, M.~A., P.~D. McNicholas, and R.~Yada (2012).
\newblock Model-based classification via mixtures of multivariate t-factor
  analyzers.
\newblock {\em Communications in Statistics -- Simulation and
  Computation\/}~{\em 41\/}(4), 510--523.

\bibitem[\protect\citeauthoryear{Subedi, Punzo, Ingrassia, and
  McNicholas}{Subedi et~al.}{2013}]{subedi13}
Subedi, S., A.~Punzo, S.~Ingrassia, and P.~D. McNicholas (2013).
\newblock Clustering and classification via cluster-weighted factor analyzers.
\newblock {\em Advances in Data Analysis and Classification\/}~{\em 7\/}(1),
  5--40.

\bibitem[\protect\citeauthoryear{Subedi, Punzo, Ingrassia, and
  McNicholas}{Subedi et~al.}{2015}]{subedi15}
Subedi, S., A.~Punzo, S.~Ingrassia, and P.~D. McNicholas (2015).
\newblock Cluster-weighted t-factor analyzers for robust model-based clustering
  and dimension reduction.
\newblock {\em Statistical Methods and Applications\/}~{\em 24\/}(4), 623--649.

\bibitem[\protect\citeauthoryear{Tiedeman}{Tiedeman}{1955}]{tiedeman55}
Tiedeman, D.~V. (1955).
\newblock On the study of types.
\newblock In S.~B. Sells (Ed.), {\em Symposium on Pattern Analysis}. Randolph
  Field, Texas: Air University, U.S.A.F. School of Aviation Medicine.

\bibitem[\protect\citeauthoryear{Tortora, Browne, Franczak, and
  McNicholas}{Tortora et~al.}{2015}]{tortora15c}
Tortora, C., R.~P. Browne, B.~C. Franczak, and P.~D. McNicholas (2015).
\newblock {\em MixGHD: Model Based Clustering, Classification and Discriminant
  Analysis Using the Mixture of Generalized Hyperbolic Distributions}.
\newblock R package version 1.8.

\bibitem[\protect\citeauthoryear{Venables and Ripley}{Venables and
  Ripley}{2002}]{MASS}
Venables, W.~N. and B.~D. Ripley (2002).
\newblock {\em Modern Applied Statistics with S\/} (Fourth ed.).
\newblock New York: Springer.
\newblock ISBN 0-387-95457-0.

\bibitem[\protect\citeauthoryear{Vrbik and McNicholas}{Vrbik and
  McNicholas}{2012}]{vrbik12}
Vrbik, I. and P.~D. McNicholas (2012).
\newblock Analytic calculations for the {EM} algorithm for multivariate skew-t
  mixture models.
\newblock {\em Statistics and Probability Letters\/}~{\em 82\/}(6), 1169--1174.

\bibitem[\protect\citeauthoryear{Vrbik and McNicholas}{Vrbik and
  McNicholas}{2014}]{vrbik14}
Vrbik, I. and P.~D. McNicholas (2014).
\newblock Parsimonious skew mixture models for model-based clustering and
  classification.
\newblock {\em Computational Statistics and Data Analysis\/}~{\em 71},
  196--210.

\bibitem[\protect\citeauthoryear{Vrbik and McNicholas}{Vrbik and
  McNicholas}{2015}]{vrbik15}
Vrbik, I. and P.~D. McNicholas (2015).
\newblock Fractionally-supervised classification.
\newblock {\em Journal of Classification\/}~{\em 32\/}(3), 359--381.

\bibitem[\protect\citeauthoryear{Wolfe}{Wolfe}{1965}]{wolfe65}
Wolfe, J.~H. (1965).
\newblock A computer program for the maximum likelihood analysis of types.
\newblock Technical Bulletin 65-15, U.S.\ Naval Personnel Research Activity.

\end{thebibliography}

\clearpage
\appendix
\section{tEIGEN Models}\label{app:tEIGEN}
\begin{table}[!htb]
\centering
\caption{Model nomenclature and number of free covariance parameters of tEIGEN models with constrained (C), unconstrained (U) and identity (I) elements.}
\begin{tabular}{llllll}
\hline
Model& $\lambda_g=\lambda$ & ${\boldsymbol \Lambda}_g={\boldsymbol \Lambda}$& ${\bf D}_g={\bf D}$& $\nu_g=\nu$ & No. of Free Covariance Parameters\\
\hline
CIIC & C&I&I&C&1+1\\
CIIU &C&I&I&U&$1+G$\\
UIIC &U&I&I&C&$(G-1)+1$\\
UIIU&U&I&I&U&$(G-1)+G$\\
CICC&C&I&C&C&$p+1$\\
CICU&C&I&C&U&$p+G$\\
UICC&U&U&C&C&$p+(G-1)+1$\\
UICU&U&I&C&U&$p+(G-1)+G$\\
CIUC&C&I&U&C&$Gp-(G-1)+1$\\
CIUU&C&I&U&U&$Gp-(G-1)+G$\\
UIUC&U&I&U&C&$Gp+1$\\
UIUU&U&I&U&U&$Gp+G$\\
CCCC&C&C&C&C&[$p(p+1)/2]+1$\\
CCCU&C&C&C&U&[$p(p+1)/2]+G$\\
UCCC&U&C&C&C&[$p(p+1)/2]+(G-1)+1$\\
UCCU&U&C&C&U&[$p(p+1)/2]+(G-1)+G$\\
CUCC&C&U&C&C&$G[p(p+1)/2]-(G-1)(p)+1$\\
CUCU&C&U&C&U&$G[p(p+1)/2]-(G-1)(p)+G$\\
UUCC&U&U&C&C&$G[p(p+1)/2]-(G-1)(p-1)+1$\\
UUCU&U&U&C&U&$G[p(p+1)/2]-(G-1)(p-1)+G$\\
CCUC&C&C&U&C&[$p(p+1)/2]+(G-1)(p-1)+1$\\
CCUU&C&C&U&U&[$p(p+1)/2]+(G-1)(p-1)+G$\\
CUUC&C&U&U&C&$G[p(p+1)/2]-(G-1)+1$\\
CUUU&C&U&U&U&$G[p(p+1)/2]-(G-1)+G$\\
UCUC&U&C&U&C&$G[p(p+1)/2]+(G-1)p+1$\\
UCUU&U&C&U&U&$G[p(p+1)/2]+(G-1)p+G$\\
UUUC&U&U&U&C&$G[p(p+1)/2]+1$\\
UUUU&U&U&U&U&$G[p(p+1)/2]+G$\\
\hline
\end{tabular}
\label{tab:teigen}
\end{table}

\section{Alternative Form of the Likelihood}\label{app:b}
\subsection{Alternative Likelihood}
We have already seen that the observed weighted likelihood can be written as in \eqref{eq:orig_likelihood} and, analogous to \eqref{eq:cwll}, the associated complete-data weighted likelihood can be written as
\begin{equation}
\mL\tsub{\tiny comp}(\btheta|D\tsub{C},\alpha)=\prod_{i=1}^2\left[\prod_{j=1}^{n_i}\prod_{g=1}^{G}[\pig f_g(\bfx_{ij}|\btheta)]^{z_{jg}^{(i)}}\right]^{\alpha_i}.
\label{eq:CWL}
\end{equation}
\cite{dempster77} state that when integrating the complete-data likelihood over the space of unknown quantities, in our case $\mathbb{Z}_2$, it is desired that the result should be the observed likelihood. The observed likelihood as given in \eqref{eq:orig_likelihood}, however, does not satisfy this property. Indeed,
\begin{align}
\int_{\mathbb{Z}_2}\mL\tsub{\tiny comp}(\bvtheta|D\tsub{C},\alpha)d{\bf z}_{2} &= \int_{\mathbb{Z}_2}\left\{\prod_{j=1}^{n_1}\prod_{g=1}^{G}[\pig f_g(\bfx_{1j}|\btheta_g)]^{z_{jg}^{(1)}\alpha}\times\prod_{j=1}^{n_2}\prod_{g=1}^{G}\left[\pig f_g(\bfx_{2j}|\btheta_g)\right]^{z_{jg}^{(2)}(1-\alpha)}\right\}d{\bf z}_2 \nonumber \\
&=\prod_{j=1}^{n_1}\prod_{g=1}^{G}[\pig f_g(\bfx_{1j}|\btheta_g)]^{z_{jg}^{(1)}\alpha}\prod_{j=1}^{n_2}\left\{\int_{\mathbb{Z}_2}\prod_{g=1}^{G}\left[\pig f_g(\bfx_{2j}|\btheta_g)\right]^{z_{jg}^{(2)}(1-\alpha)} d\bfz_2\right\} \nonumber \\
&=\prod_{j=1}^{n_1}\prod_{g=1}^{G}[\pig f_g(\bfx_{1j}|\btheta)]^{z_{jg}^{(1)}\alpha} \prod_{j=1}^{n_2}\left\{\sum_{{\bf z}_{j}\in\mathfrak{B}} \prod_{g=1}^{G}\left[\pig f_g(\bfx_{2j}|\btheta)\right]^{z_{jg}^{(2)}(1-\alpha)}\right\} \nonumber \\
&=\prod_{j=1}^{n_1}\prod_{g=1}^{G}[\pig f_g(\bfx_{1j}|\btheta)]^{z_{jg}^{(1)}\alpha}\prod_{j=1}^{n_2}\left\{\sum_{g=1}^{G}\left[\pig f_g(\bfx_{2j}|\btheta)\right]^{(1-\alpha)}\right\}, \label{eq:alt_likelihood}
\end{align} 
where 
$$
\mathfrak{B}=\left\{{\bf z}_{j}=\left(z_{j1}^{(2)},z_{j2}^{(2)},\ldots,z_{jG}^{(2)} \right)~\Big|~z_{jg}^{(2)}\in\{0,1\}, \forall g \in\{1,2,\ldots,G\},\sum_{g=1}^{G}z_{jg}^{(2)}=1\right\}.
$$
Clearly, this is not the same as the form given in \eqref{eq:orig_likelihood}. Therefore, to maintain the relationship between the complete and incomplete weighted likelihood as presented in \cite{dempster77}, we consider using the form of the incomplete weighted likelihood given in \eqref{eq:alt_likelihood} and denote this by $\mathcal{L}\tsub{\tiny alt}$.

Note that there are two extreme cases that should be considered separately. The first extreme case is when $\alpha=0$:
\begin{align*}
\int_{\mathbb{Z}_2}\mathcal{L}\tsub{\tiny comp}(\bvtheta|D\tsub{C},\alpha=0)d\bfz_2&=\int_{\mathbb{Z}_2}\prod_{j=1}^{n_2}\prod_{g=1}^{G}\left[\pig f_g(\bfx_{2j}|\btheta_g)\right]^{z_{jg}^{(2)}(1-\alpha)}d{\bf z}_2=\prod_{j=1}^{n_2}\sum_{g=1}^{G}\pig f_g(\bfx_{2j}|\btheta),
\end{align*}
which is equivalent to \eqref{eq:alt_likelihood} when $\alpha=0$.
The second extreme case, which turns out to be more interesting, is when $\alpha=1$:
\begin{align*}
\int_{\mathbb{Z}_2}\mathcal{L}\tsub{\tiny comp}(\bvtheta|D\tsub{C},\alpha=1)d\bfz_2&=\int_{\mathbb{Z}_2}\prod_{j=1}^{n_1}\prod_{g=1}^{G}[\pig f_g(\bfx_{1j}|\btheta_g)]^{z_{jg}^{(1)}}d\bfz_2
=\prod_{j=1}^{n_1}\prod_{g=1}^{G}[\pig f_g(\bfx_{1j}|\btheta_g)]^{z_{jg}^{(1)}},
\end{align*}
which is the same as $\mathcal{L}\tsub{\tiny DA}$, the observed likelihood for a discriminant analysis. However, in \eqref{eq:alt_likelihood}, when $\alpha=1$, 
\begin{align*}
\mathcal{L}\tsub{\tiny alt}(\bvtheta|D\tsub{\tiny o})&=n_2\prod_{j=1}^{n_1}\prod_{g=1}^{G}[\pig f_g(\bfx_{1j}|\btheta_g)]^{z_{jg}^{(1)}}=n_2\mathcal{L}\tsub{\tiny DA}(\bvtheta|D\tsub{\tiny L}) \numberthis \label{eq:alt_disc}.
\end{align*}
When $\alpha=1$ we are performing a discriminant analysis and so the form of the observed and weighted likelihoods should be the same, which is clearly not the case. Therefore, when $\alpha=1$, we use $\mathcal{L}\tsub{\tiny DA}$ for our observed likelihood.

For both the original and altered observed likelihoods, the complete-data likelihood is identical. Therefore, if we were to take a Gaussian mixture model, the updates in the M-step would be the same as those given in \cite{vrbik15}, regardless of whether the original or alternative likelihood were used. However, the updates for $\hat{z}_{jg}^{(2)}$ in the E-step would become
\begin{align*}
\hat{z}_{jg}^{(2)}=\frac{\left[\pig^{(t)}\phi(\bfx_{2j}|\mug^{(t)},\Sigg^{(t)})\right]^{(1-\alpha)}}{\sum \limits_{g=1}^{G}\left[\pig^{(t)}\phi(\bfx_{2j}|\mug^{(t)},\Sigg^{(t)})\right]^{(1-\alpha)}}.
\end{align*}

\subsection{Simulation Comparing the Original and Altered Likelihoods} \label{sims:origAlt}
We perform simulations to compare the performance of the original and altered likelihoods. We simulate 100 datasets with 300 samples: 150 of these sample belong to one group which follows a $\mathcal{N}_2({\bf 0},{\mathbf \Sigma}_1)$, and the remaining 150 belong to another group which follows a $\mathcal{N}_2({\mathbf \Delta},{\mathbf \Sigma}_2)$, where ${\mathbf \Delta}=\left[0,\Delta\right]',$ and
\begin{align*}
{\mathbf \Sigma}_1=
\left[\begin{array}{cc}
1&0.7\\
0.7&1\\
\end{array}\right],
& \hspace{1cm}
{\mathbf \Sigma}_2=
\left[\begin{array}{cc}
1&0\\
0&1\\
\end{array}\right].
\end{align*}
We take $\Delta\in\{1,5\}$ corresponding to different levels of clustering difficulty. For each dataset, we consider $p\in\{10,20,\ldots,80,90\}$, where $p$ is the percentage of labelled data. 

To choose the weights for FSC, we looked at 11 different values of $\alpha$. These values were taken to be $\alpha\in{\boldsymbol \alpha}_{\tiny\mbox{ARI}}$, where ${\boldsymbol \alpha}_{\tiny\mbox{ARI}}=\{0,0.1,0.2,\ldots,1\}$. We then calculate the ARI for each of these weights for the 100 datasets and take the average ARI for each weight. We then choose the weight that had the highest average ARI. We denote the resulting FSC solution for each weight $\alpha$ by FSC\tsub{$\alpha$}. Furthermore, denote by FSC\tsub{\tiny ARI} the FSC solution with the chosen weight resulting from the highest average ARI. Finally, in the special cases corresponding to the three species of classification $\alpha=0,0.5,1$, we denote the FSC solution by FSC\tsub{\tiny clust}, FSC\tsub{\tiny class} and FSC\tsub{\tiny DA}, respectively. 

In Figures \ref{fig:avod1} and \ref{fig:avod5}, we show different line plots for $\Delta=1$ and $\Delta=5$, respectively. In each plot, the average ARI is plotted against the percentage of labelled data $p$. A dotted black line is used to show the result for FSC\tsub{\tiny ARI} with the corresponding chosen weight shown above each point. The first row in each plot shows the results when using all the weights, and the second row singles out the three different species of classification and FSC\tsub{\tiny ARI} The standard errors were calculated by taking the ARI for all 100 datasets of the chosen weight of FSC\tsub{\tiny ARI} and calculating one (darker grey) and two (lighter grey) standard deviations from the mean ARI. 
\begin{figure}[!htb]
\centering
\includegraphics[width=0.85\textwidth]{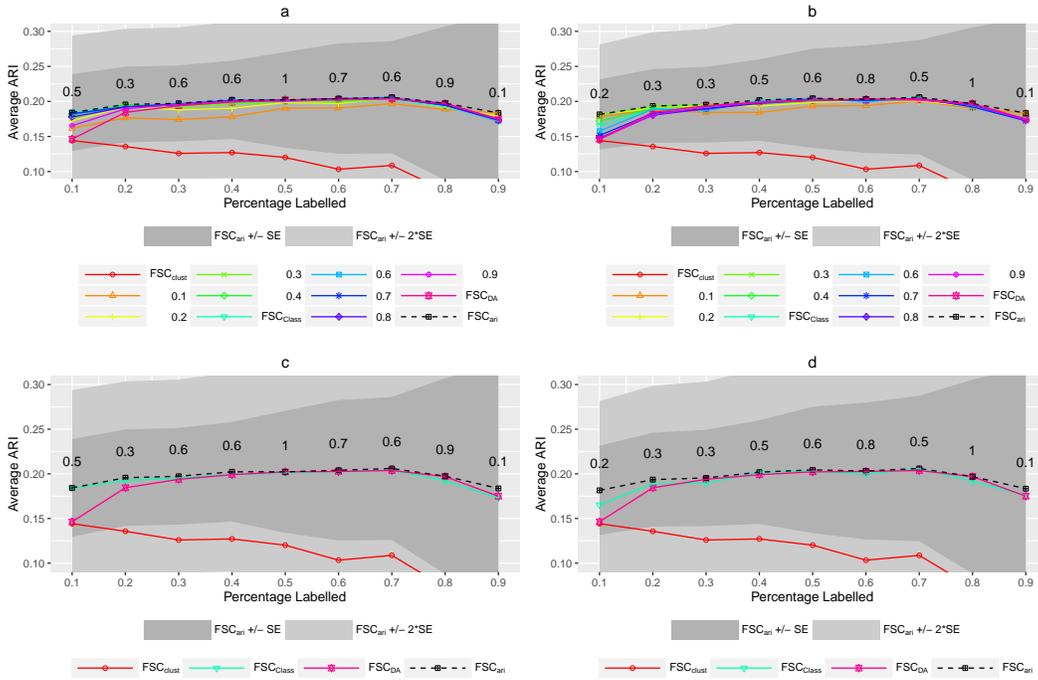}
\caption[Altered vs Original Likelihood Results with $\Delta=1$]{For $\Delta=1$: (a) and (b) FSC\tsub{$\alpha$} and FSC\tsub{\tiny ARI} ($\alpha\in\alphset$) for the original and altered likelihood respectively. (c) and (d) FSC\tsub{\tiny clust}, FSC\tsub{\tiny class}, FSC\tsub{\tiny DA} and FSC\tsub{\tiny ARI} for the original and altered likelihood respectively.}
\label{fig:avod1}
\end{figure}
\begin{figure}[!ht]
\centering
\includegraphics[width=0.85\textwidth]{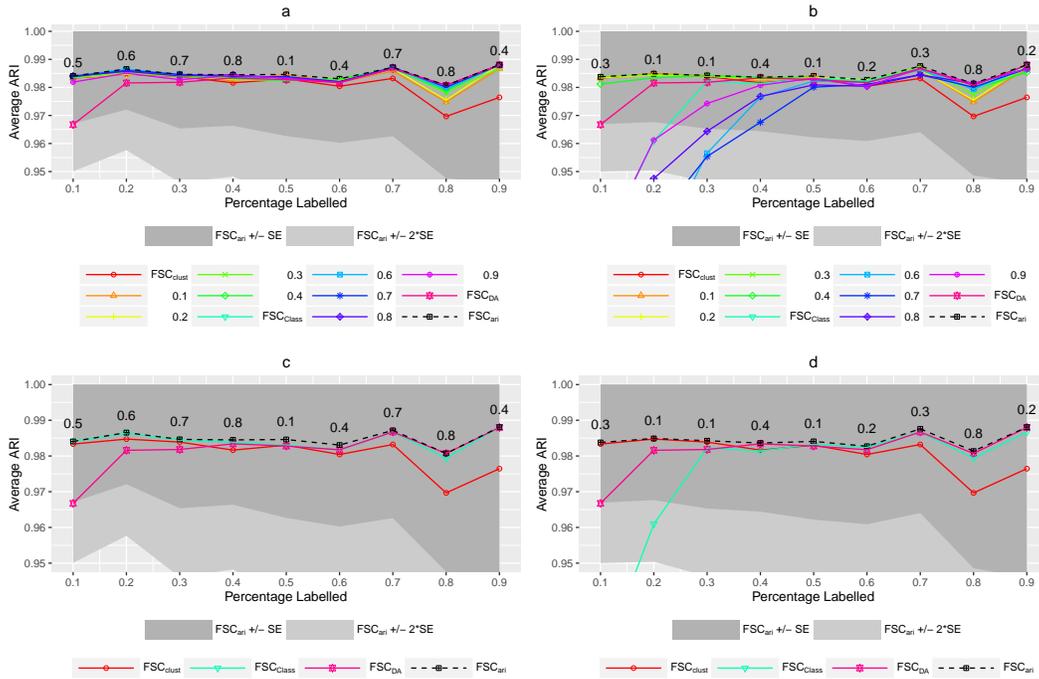}
\caption[Altered vs Original Likelihood Results with $\Delta=5$]{For $\Delta=5$: (a) and (b) FSC\tsub{$\alpha$} and FSC\tsub{\tiny ARI} ($\alpha\in\alphset$) for the original and altered likelihood respectively. (c) and (d) FSC\tsub{\tiny clust}, FSC\tsub{\tiny class}, FSC\tsub{\tiny DA} and FSC\tsub{\tiny ARI} for the original and altered likelihood respectively.}
\label{fig:avod5}
\end{figure}

In general, the overall classification performance between the altered and original likelihoods are similar. The chosen weights for FSC\tsub{\tiny ARI}, however, differ between the two forms of the likelihood. For $\Delta=1$, this difference is less pronounced than for $\Delta=5$. More specifically, for $\Delta=1$, the difference between the weights for all but 10\%, 30\% and 50\% differ by at most 0.1 if they are not exactly the same. For $\Delta=5$, however, the differences between the chosen weights are greater, and there are fewer proportions for which the difference is small. We also see that at lower percentages of labelled data, there is more variability in the average ARI between the different weights.
In conclusion, although the choice of the weights are different between the two likelihoods, the overall classification performance when using the chosen weight in each case are very similar. Moreover, the altered form is not strictly a likelihood. Accordingly, we will henceforth use the original, and more natural form, form of the likelihood for FSC.

\end{document}